\documentclass[twocolumn,Astrosymb]{aastex631}

\usepackage{kotex}
\usepackage{verbatim}
\usepackage{xcolor}
\usepackage{graphicx} 
\usepackage{multirow}
\usepackage{float}
\usepackage{natbib}
\usepackage[version=4]{mhchem} 

\newcommand{\imagescaleOne}{0.3} 

\begin{document}
\title{Precision Analysis of $\mathrm{^{12}C / ^{13}C}$ Ratios in Orion IRc2 Acetylene Isotopologues via $\chi^2$ Fitting}

\author{Minkyu Lee}
\affiliation{Department of Physics and Origin of Matter and Evolution of Galaxies (OMEG) Institute, Soongsil University, Seoul 06978, Republic of Korea}

\author{Jubin Park}
\affiliation{Department of Physics and Origin of Matter and Evolution of Galaxies (OMEG) Institute, Soongsil University, Seoul 06978, Republic of Korea}

\author{Sehoon Oh}
\affiliation{Department of Physics and Origin of Matter and Evolution of Galaxies (OMEG) Institute, Soongsil University, Seoul 06978, Republic of Korea}

\author{Myung-Ki Cheoun}
\affiliation{Department of Physics and Origin of Matter and Evolution of Galaxies (OMEG) Institute, Soongsil University, Seoul 06978, Republic of Korea}

\author{Se Young Park}
\affiliation{Department of Physics and Origin of Matter and Evolution of Galaxies (OMEG) Institute, Soongsil University, Seoul 06978, Republic of Korea}

\begin{abstract}
We present a detailed analysis of acetylene (C$_2$H$_2$) and its isotopologues in the Orion IRc2 region, focusing on the determination of $^{12}$C/$^{13}$C isotopic ratios using high-resolution infrared spectra from SOFIA. By employing a robust $\chi^2$ fitting method, we simultaneously determined temperature and column density, achieving a $^{12}$C/$^{13}$C ratio of $18.72^{+1.54}_{-1.46}$ for the blue clump and $15.07^{+1.61}_{-1.60}$ for the red clump. These results revealed significant discrepancies with the traditional rotational diagram method, which overestimated the ratios by 12.1\% and 23.9\%, respectively. Our $\chi^2$ approach also reduced uncertainties by up to 
75\% 
, providing more precise and reliable isotopic ratios. Additionally, we extended the analysis to isotopologues not covered in HITRAN, calculating vibrational and rotational constants through quantum chemical calculations. This allowed us to model subtle isotopic shifts induced by $^{13}$C and deuterium substitution, enabling accurate isotopologue detection in astrophysical environments. The Python package (TOPSEGI) developed in this study facilitates efficient $\chi^2$ fitting and isotopic ratio analysis, making it a valuable tool for future high-resolution observations. This work highlights the critical role of advanced spectral models and fitting techniques in understanding isotopic fractionation and the chemical evolution of interstellar matter.
\end{abstract}

\section{Introduction}
The study of ro-vibrational transitions in simple molecules such as acetylene (C$_2$H$_2$) \footnote{Hereafter, unless otherwise specified, the element C refers exclusively to $^{12}$C. For carbon isotopes \citep{woods2009carbon}, $^{13}$C is explicitly indicated whenever it is mentioned.} 
and its isotopologues plays a crucial role in understanding the chemical evolution of interstellar environments. Acetylene, a widely observed molecule in planetary atmospheres, interstellar clouds, and carbon-rich stars (see Table \ref{tab:occurrence_of_acetylene}\citep{pentsak2024role}), exhibits distinct infrared (IR) spectra that are sensitive to isotope substitution. Among these isotopologues, $^{13}$C-substituted acetylene ($^{13}$CCH$_2$) is of particular interest, as its detection can provide valuable insights into the isotopic composition of interstellar matter and the processes driving galaxy chemical evolution \citep{jacob2020first}.

Carbon, the primary element in acetylene and its isotopologues, is synthesized mainly in intermediate-mass stars (approximately 0.8M$_\odot$ to 8M$_\odot$) through thermonuclear fusion reactions \citep{karakas2010updated}. $^{12}$C is the dominant product of the triple-alpha process, while $^{13}$C is produced over longer timescales as a secondary product of stellar nucleosynthesis. In asymptotic giant branch (AGB) stars, $^{13}$C is primarily formed as a byproduct of the CNO cycle. This process begins with the fusion of $^{12}$C and protons, resulting in $^{13}$N, which subsequently decays via positron emission into $^{13}$C \citep{pagel2009nucleosynthesis}. Once formed, carbon combines into simple carbon-bearing molecules such as acetylene (C$_2$H$_2$), CO and HCN within stellar atmospheres \citep{mccarthy2019building}. These molecules are then dispersed into the interstellar medium (ISM) by stellar winds, enabling the formation of additional carbon compounds. The widespread distribution of these carbon-bearing molecules across the universe provides a unique opportunity to trace stellar nucleosynthesis processes and the chemical evolution of interstellar environments.

\twocolumngrid
\begin{deluxetable*}{lllllll}
\tablewidth{0pt}
\tablecaption{Occurrence of acetylene in various astrophysical environments and objects. The table summarizes the detection of $\mathrm{C_2H_2}$ across different objects, including planets, satellites, comets, and stars, along with the Orion IRc2 region. Columns provide details on the object type, measurement technique used (e.g., ISO-SWS, TEXES, SOFIA/EXES), year of investigation, spectral range in wavenumber (cm$^{-1}$) and wavelength ($\mu$m), and the corresponding references. This data highlights the widespread presence of acetylene and the variety of observational tools employed to study it in diverse cosmic settings \citep{pentsak2024role}. Among these, we focus on the frequency ranges exhibiting infrared activity in various acetylene observation data.} \label{tab:occurrence_of_acetylene}
\tablehead{
\multicolumn{2}{c}{Object} & Measurement technique & Investigation & \multicolumn{2}{c}{Range}              & Ref. \\
&                          &                       & Year                      & $\mathrm{cm^{-1}}$ & $ \mu \mathrm{m}$ & 
}
\startdata
\textbf{Planets} & Jupiter & Cassini/CIRS mid infrared spectra                                & 2006 & 670 $\sim$ 760     & 14.9 $\sim$ 13.2 & (1)  \\
                 & Saturn  & ISO-SWS                                                          & 1997 & 666.7 $\sim$ 714.3 & 15   $\sim$ 14   & (2)          \\
                 & Uranus  & ISO (Infrared Space Observatory)                                 & 1998 & 694.4 $\sim$ 769.2 & 13   $\sim$ 14.4 & (3) \\
                 & Neptune & Voyager 2/IRIS                                                   & 1991 & 720 $\sim$ 740     & 13.9 $\sim$ 13.5 & (4)\\
\hline
\textbf{Satellites} & Titan    & TEXES                                                     & 2017 & 742.9 $\sim$ 746.7 & 13.5 $\sim$ 13.4 & (5) \\
\hline
\textbf{Comets} & Hyakutake & Infrared Telescope Facility at Mauna Kea                        & 1996 & 3282 $\sim$ 3288   & 3.047 $\sim$ 3.041 & (6) \\
                &           & Cryogenic echelle spectrometer (CSHELL)  &      & & \\
\hline
\textbf{Carbon star} & IRC +10216 & TEXES &  2008 & 714.29 $\sim$ 909.09 & 14 $\sim$ 11 & (7)\\
\hline
\textbf{Orion IRc2}          &  & SOFIA/EXES                                                   & 2018 & 750.19 $\sim$ 771.60 & 13.33 $\sim$ 12.96 & (8)\\
\enddata

\tablerefs{
(1) \citealt{nixon2007meridional};
(2) \citealt{de1997first};
(3) \citealt{encrenaz1998iso};
(4) \citealt{bezard1991hydrocarbons};
(5) \citealt{bezard2022d};
(6) \citealt{brooke1996detection};
(7) \citealt{fonfria2007detailed};
(8) \citealt{rangwala2018high};
}
\end{deluxetable*}

The $^{12}$C/$^{13}$C ratio, determined through the study of carbon-bearing molecules such as acetylene (C$_2$H$_2$), serves as an essential diagnostic tool for tracing the history of nucleosynthesis and the chemical evolution of galaxies. In the solar system, this ratio is approximately 80 \citep{woods2009carbon}, whereas in carbon stars like Y CVn, it has been observed to be as low as 3.5 \citep{lambert1986chemical}, indicating a significant enhancement of $^{13}$C in certain astrophysical environments. These variations in isotopic ratios offer crucial insights into processes such as stellar nucleosynthesis, mass loss in evolved stars, and the chemical enrichment of the interstellar medium (ISM). By analyzing ro-vibrational spectra of molecules like acetylene, precise isotopic ratios can be derived, enabling a deeper understanding of the chemical and physical conditions within diverse astrophysical environments.

Among various carbon-bearing molecules, acetylene (C$_2$H$_2$) and its isotopologues offer a particularly effective means for probing $^{12}$C/$^{13}$C ratios in interstellar environments. Infrared (IR) spectroscopy of acetylene enables the precise estimation of molecular column densities ($N$) by leveraging rotational quantum numbers ($J$) and total internal partition sums (TIPS) under the assumption of local thermodynamic equilibrium (LTE). These molecular abundances, when combined with isotopologue-specific spectra, provide valuable information on the chemical and physical conditions of interstellar clouds and star-forming regions. However, traditional analysis methods, such as the rotational diagram (RD) method \citep{goldsmith1999population}, often overestimate $^{12}$C/$^{13}$C ratios due to their reliance on linear extrapolations and the separate fitting of temperature ($T$) and column density. These limitations introduce systematic biases and reduce the reliability of the derived isotopic ratios, highlighting the need for more robust and accurate fitting approaches, such as the $\chi^2$ method proposed in this study.

\begin{figure}
\includegraphics[width=\columnwidth]{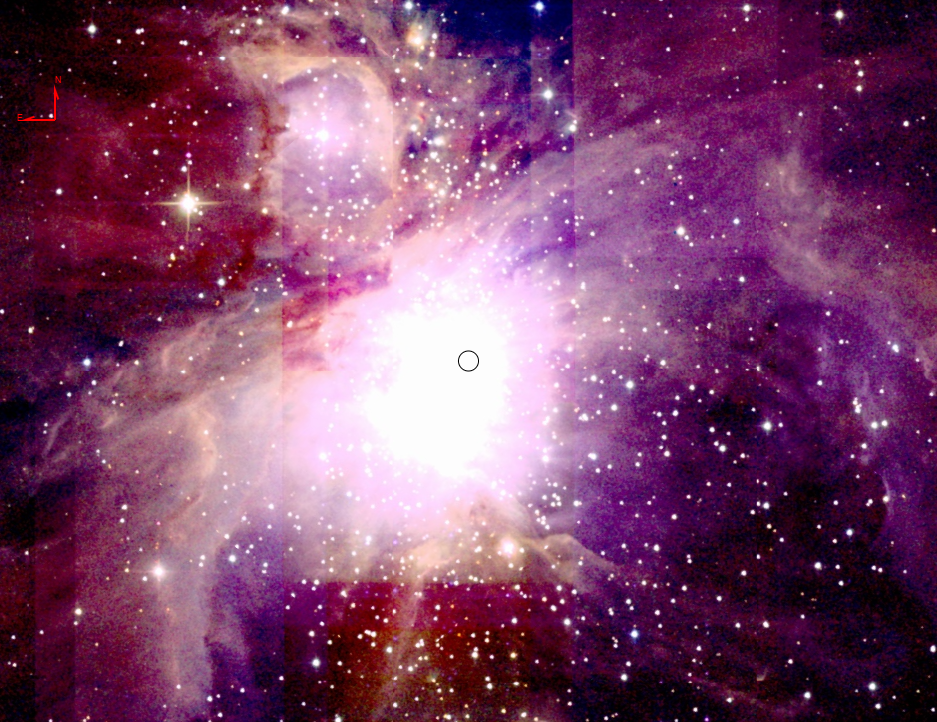}\\\
\caption{
The black circle at the center of the figure marks the location of the Orion IRc2 region, within the Becklin-Neugebauer/Kleinmann-Low (BN/KL) complex of the Orion Molecular Cloud (OMC-1) \citep{HiPS}. This region, highlighted by its dense interstellar clouds and illuminated cosmic dust, serves as a chemically rich environment for studying isotopic fractionation processes. The bright core represents the IRc2 region, where our analysis focuses on the distinct kinematic components: the blue clump ($v_\mathrm{LSR} = -7.1 \pm 0.7$ km s$^{-1}$) and the red clump ($v_\mathrm{LSR} = 1.4 \pm 0.5$ km s$^{-1}$) \citep{nickerson2023mid}. These components exhibit unique molecular compositions, providing an ideal testbed for comparing the $\chi^2$ fitting method with traditional rotational diagram (RD) approaches.}\label{fig:Orion_IRC2}
\end{figure}

To overcome the limitations of traditional methods, we implemented the $\chi^2$ fitting approach to analyze the Orion IRc2 region, a chemically diverse and dynamically intricate site of massive star formation within the Becklin-Neugebauer/Kleinmann-Low (BN/KL) region of the Orion Molecular Cloud (OMC-1). This area, illustrated in Figure \ref{fig:Orion_IRC2}, features bright stars embedded within dense interstellar clouds illuminated by cosmic dust. Such a complex and rich environment provides an ideal setting for studying isotopic fractionation processes and their role in the chemical evolution of interstellar matter \citep{stahler2008formation}. Specifically, our analysis targeted the kinematic components of the IRc2 region, referred to as the blue and red clumps, which exhibit distinct line-of-sight velocities ($v_\mathrm{LSR} = -7.1 \pm 0.7$~km~s$^{-1}$ for the blue clump and $v_\mathrm{LSR} = 1.4 \pm 0.5$~km~s$^{-1}$ for the red clump \citep{nickerson2023mid}) and unique molecular compositions. The high spectral resolution and dynamic range of this region make it particularly suitable for evaluating the accuracy and advantages of the $\chi^2$ method in isotopic ratio determinations.

Using high-resolution infrared spectra from SOFIA/EXES, the $\chi^2$ fitting method determined temperature ($T$) and column density ($N$) simultaneously by minimizing residuals between observed and theoretical lower-state column densities. Compared to the rotational diagram (RD) method, the $\chi^2$ approach reduced uncertainties by up to 
75\%
in the blue clump and corrected RD overestimations of $^{12}\mathrm{C}/^{13}\mathrm{C}$ ratios by 12.1\% and 23.9\% for the blue and red clumps, respectively. These results underscore the utility of advanced techniques like the $\chi^2$ method in probing isotopic and chemical properties of massive star-forming regions. Furthermore, the spatial and kinematic separation of the blue and red clumps reveals their distinct physical origins and evolutionary processes within the BN/KL complex.

To improve efficiency and broaden the scope of our analysis, we incorporated a QuadTree search algorithm into the $\chi^2$ fitting method, enabling targeted exploration of parameter space and minimizing computational overhead. Furthermore, we extended our study to all isotopologues of acetylene substituted with $^{13}$C and deuterium (D), calculating their rotational and vibrational constants using quantum chemical methods such as Density Functional Theory (DFT) and Møller-Plesset perturbation theory (MP2). These calculations filled critical gaps in existing data, particularly for isotopologues absent in HITRAN \citep{JACQUEMART2003363, gordon2022hitran2020}, and revealed significant isotopic effects, including shifts in vibrational modes caused by $^{13}$C and D substitutions.

By combining advanced computational techniques with high-resolution observations, our approach offers a robust framework for precise isotopic ratio determinations. The Python package developed in this study further strengthens this framework by automating the analysis and enabling direct comparisons between the RD and $\chi^2$ methods. This tool simplifies isotopic ratio calculations, facilitating future research into isotopic fractionation, molecular abundances, and the chemical evolution of diverse astrophysical environments with greater precision and reduced biases.

The structure of this paper is as follows. Section~\ref{sec:observations} provides an overview of the observational data, including previous spectral analyses of acetylene and its isotopologues such as $^{13}$CCH$_2$ and C$_2$HD. Section~\ref{sec:results} presents the findings of our reanalysis in three parts: Section~\ref{subsec:results_isotopologues} focuses on the results for C$_2$H$_2$, $^{13}$CCH$_2$, and C$_2$HD; Section~\ref{subsec:results_theoretical} discusses theoretical calculations for seven isotopologues absent from the HITRAN database; and Section~\ref{subsec:ratio_comparison} compares the $\mathrm{^{12}C / ^{13}C}$ isotopic ratios derived using the RD and $\chi^2$ methods,
summarizing the differences and improvements. Finally, Section~\ref{sec:conclusions} concludes the paper by highlighting the implications of these findings, particularly the advantages of the $\chi^2$ fitting method for isotopic studies in astrophysical environments.

Details of the quantum chemical tools used, including Gaussian16 and DFT calculations, are summarized in the Appendix. The Appendix also provides precomputed spectral data, instructions for using the Python package developed in this study, and guidance on accessing the code via GitHub. This package enables automation of the fitting process, offering a direct comparison between RD and $\chi^2$ methods while filling critical gaps in existing datasets, thereby supporting future research in isotopic ratio determinations.

\section{Observations}\label{sec:observations}
\subsection{A Brief Introduction to Acetylene Molecules}
\twocolumngrid
\begin{deluxetable*}{ccl}
\tablecaption{Normal vibrational modes of C$_2$H$_2$. The table lists the five fundamental modes: symmetric and anti-symmetric stretching and bending. The $\nu_1$,  $\nu_4$ and $\nu_5$ modes are infrared inactive due to their symmetry, while the $\nu_3$ and $\nu_5$ modes are infrared active. The bending modes, $\nu_4$ and $\nu_5$, are doubly degenerate, contributing to the overall vibrational spectrum of the molecule \citep{chubb2018rotation}.}
\label{tab:normal_modes}
\tablehead{
    \colhead{Label} & \colhead{Normal mode} & \colhead{Description}
}
\startdata
\raisebox{0.5\height}{$\nu_{1}$} &
\begin{minipage}{\imagescaleOne\textwidth}
\scalebox{\imagescaleOne}{\includegraphics{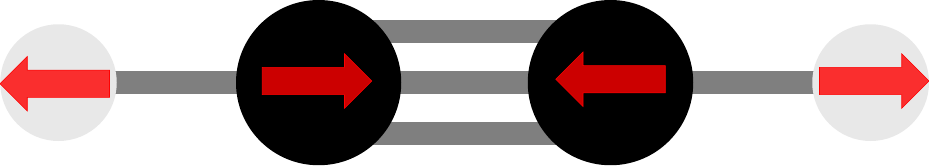}}
\end{minipage} &
$\mathrm{\mathbf{CH}}$ symmetric stretching and infrared inactivity \\
\raisebox{0.5\height}{$\nu_{2}$} &
\begin{minipage}{\imagescaleOne\textwidth}
\scalebox{\imagescaleOne}{\includegraphics{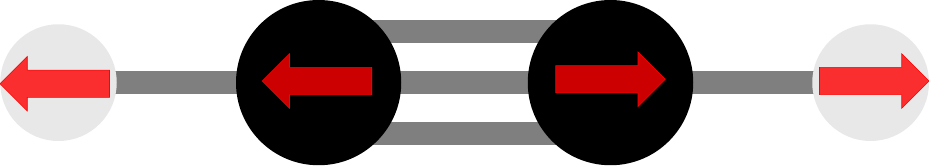}}
\end{minipage} &
$\mathrm{\mathbf{CC}}$ symmetric stretching and infrared inactivity \\
\raisebox{0.5\height}{$\nu_{3}$} &
\begin{minipage}{\imagescaleOne\textwidth}
\scalebox{\imagescaleOne}{\includegraphics{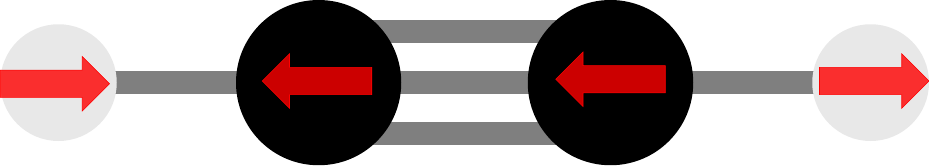}}
\end{minipage} &
$\mathrm{\mathbf{CH}}$ antisymmetric stretching and infrared activity \\
\raisebox{0.5\height}{$\nu_{4}$} &
\begin{minipage}{\imagescaleOne\textwidth}
\scalebox{\imagescaleOne}{\includegraphics{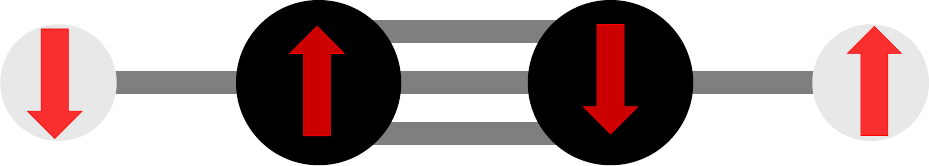}}
\end{minipage} &
Symmetric bending, infrared inactivity, and doubly degenerate  \\
\raisebox{0.5\height}{$\nu_{5}$} &
\begin{minipage}{\imagescaleOne\textwidth}
\scalebox{\imagescaleOne}{\includegraphics{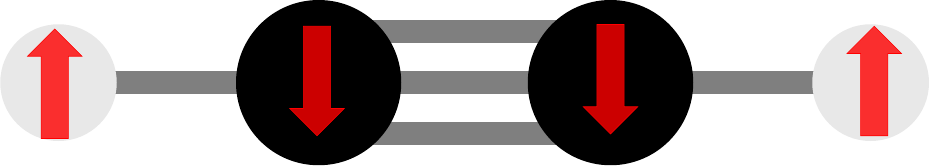}}
\end{minipage} &
Antisymmetric bending, infrared activity and doubly degenerate \\
\enddata
\end{deluxetable*}
Acetylene is a simple organic molecule widely observed in various astrophysical environments. Its lack of a permanent dipole moment prevents it from producing detectable rotational spectra in the radio range, but it is identified through ro-vibrational transitions in the mid-infrared region. Acetylene exhibits seven normal vibrational modes, characterized by symmetric and anti-symmetric stretching and bending motions. Among these, the symmetric and anti-symmetric bending modes ($\nu_4$, $\nu_5$) are doubly degenerate, reducing the number of distinct modes to five \citep{mcquarrie2008quantum}. 

As summarized in Table \ref{tab:normal_modes}, the $\nu_3$ (CH anti-symmetric stretch) and $\nu_5$ (anti-symmetric bend) modes are infrared active, while the $\nu_1$ (CH symmetric stretch), $\nu_2$ (CC symmetric stretch), and $\nu_4$ (symmetric bending) modes are infrared inactive due to their symmetry. The degeneracy of $\nu_4$ and $\nu_5$ plays a critical role in shaping the vibrational spectra of acetylene, especially when analyzing its isotopologues. This classification highlights how distinct vibrational modes contribute to the molecule’s overall spectral characteristics.

For symmetric isotopologues of acetylene, such as $^{12}$C$_2$H$_2$, $^{12}$C$_2$D$_2$, and $^{13}$C$_2$H$_2$, two nuclear spin isomers—ortho and para—are present, introducing additional complexity to the spectral analysis \citep{Herman1982acetylene, Robert2007, AMYAY201180}. The ortho isomer corresponds to parallel nuclear spins, while the para isomer has antiparallel spins, resulting in distinct population distributions among rotational energy levels. These spin isomers significantly influence the ro-vibrational spectra, facilitating the identification of acetylene isotopologues in diverse astrophysical environments. This, in turn, enhances our understanding of molecular abundances and isotopic ratios, which are critical for studying the chemical evolution of interstellar regions \citep{herman2007acetylene}.

The diverse detection of acetylene across astrophysical environments builds on this spectral complexity, showcasing its utility as a molecular tracer for interstellar chemistry. Infrared spectra of acetylene have been observed on various planets, including Jupiter, Saturn, Uranus, and Neptune, through techniques such as mid-infrared spectroscopy with Cassini/CIRS \citep{nixon2007meridional} and Voyager 2/IRIS \citep{bezard1991hydrocarbons}. Beyond planetary atmospheres, acetylene has been identified in the atmosphere of Titan \citep{bezard2022d}, the comet Hyakutake \citep{brooke1996detection}, and carbon-rich environments such as IRC +10216 \citep{fonfria2007detailed} and the Orion Molecular Cloud IRc2 \citep{rangwala2018high}. These findings illustrate how acetylene serves as a valuable diagnostic tool, tracing the processes that shape the chemical evolution of stellar and planetary systems across a range of environments and stages of formation.
\begin{figure*}[t!]
  \centering
\includegraphics[width=\textwidth]{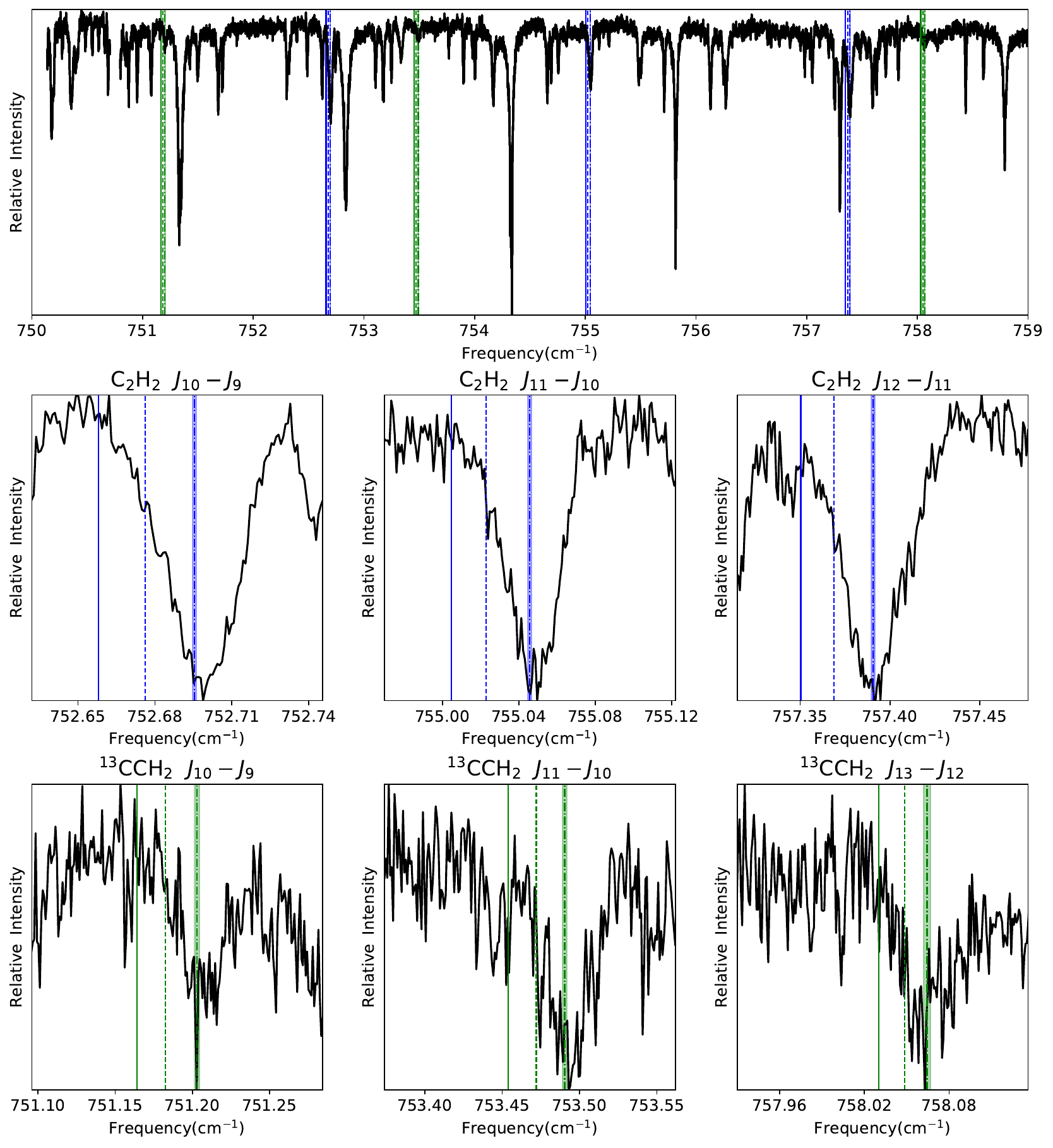}\\\
\caption{The black lines in the upper panel represent the SOFIA observational data of the IRc2 region near the Orion Nebula (M42 or NGC 1976), while the blue and green lines correspond to C$_2$H$_2$ (middle panel) and $^{13}$CCH$_2$ (lower panel), respectively. The colored solid lines are from the HITRAN (High-Resolution Transmission Molecular Absorption) database, the dashed lines indicate the Local Standard of Rest (LSR) corrections, and the dash-dotted lines show corrections for both the LSR and Doppler effects. Note that all transitions considered here correspond to the R-branch.} \label{fig:reproduce}
\end{figure*}
\subsection{SOFIA Observation Data}
Infrared (IR) observational data for acetylene and its isotopologues have been obtained using various instruments, including the Spitzer Space Telescope and the SOFIA observatory. Among these, SOFIA provides superior spectral resolution, enabling the precise resolution of individual rotational transitions that are critical for accurately determining molecular column densities \citep{rangwala2018high, nickerson2023mid}. This advantage makes SOFIA particularly well-suited for studies requiring high-resolution spectroscopic data, such as the analysis of isotopic ratios in chemically complex regions. In this study, we specifically use SOFIA observations to validate and refine our proposed methodology for isotopic ratio determination.

The mid-infrared data from SOFIA used in this study focuses on the IRc2 region near the Orion Nebula (M42 or NGC 1976), a well-known gas cloud illuminated by radiation from the Trapezium stars. Although it is neither the largest nor the brightest H II region, the Orion Nebula is one of the closest and most thoroughly studied star-forming regions \citep{o2001orion}. Approximately 0.2 pc behind the nebula are the Becklin-Neugebauer (BN) and IRc2 infrared sources, located within the Kleinman-Low (KL) nebula \citep{stahler2008formation}. The Orion BN-KL region represents one of the nearest and most extensively investigated sites of massive star formation. While much of the region's chemistry has been explored at radio, (sub)mm, and far-infrared (FIR) wavelengths, mid-infrared (MIR) spectroscopic observations, such as those of the IRc2 region, offer unique insights into its chemical composition, kinematics, geometry, and physical conditions \citep{nickerson2023mid}. This makes SOFIA’s MIR data invaluable for probing the complex environments of massive star formation.
\subsection{Spectral Analysis of $\mathrm{C_{2} H_{2}}$ and $\mathrm{^{13}CC H_{2}}$}
In molecular spectroscopy, rotational transitions are classified into three branches based on changes in the rotational quantum number $J$: the P-branch ($\Delta J = -1$), Q-branch ($\Delta J = 0$), and R-branch ($\Delta J = +1$). These transitions are fundamental to ro-vibrational spectra, revealing essential details about molecular energy states. SOFIA’s mid-infrared observations encompass all three branches, enabling a comprehensive analysis of acetylene and its isotopologues.

For acetylene (C$_2$H$_2$) and its isotopologue $^{13}$CCH$_2$, ro-vibrational transitions in the mid-infrared region, particularly around the $\nu_5$ band, allow detailed analysis of molecular properties. 
Figure~\ref{fig:reproduce} illustrates the observed spectra of C$_2$H$_2$ and $^{13}$CCH$_2$, corrected for Doppler shifts and Local Standard of Rest (LSR) velocities to ensure precise alignment with theoretical predictions from the HITRAN database. The black line represents the observed data for the Orion IRc2 region, while the blue and green lines correspond to the predicted transitions for C$_2$H$_2$ and $^{13}$CCH$_2$, respectively. The dashed lines indicate Doppler corrections, and the dash-dotted lines represent adjustments for both Doppler and LSR corrections, essential for accurately determining molecular column densities, excitation conditions, and velocity structures. Without these corrections, the spectral lines would appear shifted, leading to systematic errors. The LSR correction further ensures that the observed spectra are referenced to the standard rest frame of the interstellar medium, providing more accurate velocity measurements.

The six panels in Figure~\ref{fig:reproduce}—three above and three below the central panel—provide a comprehensive view of the ro-vibrational transitions for both ortho and para species of C$_2$H$_2$, along with the isotopologue $^{13}$CCH$_2$. The central panel displays a broader spectral range, highlighting multiple ro-vibrational transitions across the P-, Q-, and R-branches. The three panels above the central panel focus on individual transitions of C$_2$H$_2$, specifically the $J = 10 \to 9$, $J = 11 \to 10$, and $J = 12 \to 11$ transitions, with the blue vertical solid lines denoting HITRAN predictions. Similarly, the three panels below the central panel present detailed views of $^{13}$C$^{12}$CH$_2$ transitions for the same rotational quantum numbers, marked by green vertical solid lines corresponding to HITRAN predictions.

The excellent agreement between HITRAN predictions and SOFIA observations across all panels confirms the reliability of the applied corrections and the accuracy of the spectral models. These high-resolution spectra allow for precise evaluation of physical conditions within the Orion hot core, such as temperature, column density, and velocity structure, which are critical for understanding the chemical processes driving the evolution of this massive star-forming region.

\twocolumngrid
\begin{deluxetable*}{ccrrrrrl}
\tablecaption{Comparison of normal modes and rotational constants of various isotopologues of $\mathrm{C_{2} H_{2}}$ between this work (B3LYP) and related studies. The FSF values were obtained by fitting the HITRAN data. The table shows the $\nu_5$ vibrational mode and the rotational constants $\mathrm{\tilde{B}}$ for each isotopologue, highlighting the agreement between the rescaled results using FSFs from this study and previously published works. In the table, isotopic substitutions ($^{13}$C and D) are highlighted with red circles to clearly indicate the modified atoms relative to the reference molecule, $^{12}$C$_2$H$_2$.
} \label{tab:comparison_work}
\tablehead{
    \colhead{Molecule} & \colhead{Structure} & \multicolumn{3}{c|}{This work(B3LYP)} &  \multicolumn{3}{c}{Related Work} \\
    \colhead{    } & \colhead{}          & \colhead{FSF} & \colhead{$\nu_{5}$}   &  \multicolumn{1}{c|}{$\mathrm{\tilde{B}}$} & \colhead{$\nu_{5}$} & \colhead{$\mathrm{\tilde{B}}$} & \colhead{Ref.} \\ 
    \colhead{ID} & \colhead{}          &  \colhead{} &  \colhead{$\mathrm{[cm^{-1}]}$}   &  \multicolumn{1}{c|}{$\mathrm{[cm^{-1}]}$}  & \colhead{$\mathrm{[cm^{-1}]}$}  & \colhead{$\mathrm{[cm^{-1}]}$} 
}
\startdata
\multirow{2}{*}{0} &
\multirow{2}{*}{
\begin{minipage}{.3\textwidth}
\centering
\includegraphics[width=3.5cm, keepaspectratio]{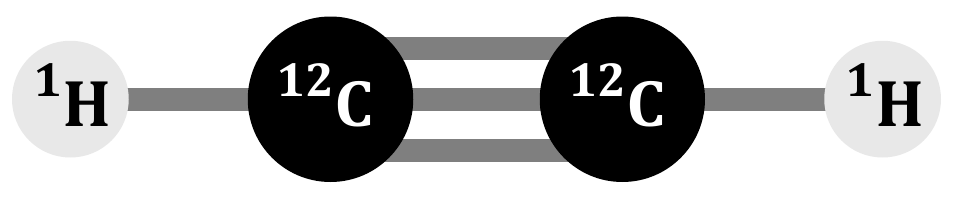}
\end{minipage}
}
  & 1.0000 & 774.627 &\multicolumn{1}{c|}{ \multirow{2}{*}{1.1766} } & \multirow{2}{*}{729.08}   & \multirow{2}{*}{1.1766}  & \multirow{2}{*}{(1), (2)}\\
& & 0.9413 & 729.185 &\multicolumn{1}{c|}{}                          &                           &                          &                                          \\ \hline
\multirow{2}{*}{1} &
\multirow{2}{*}{
\begin{minipage}{.3\textwidth}
\centering
\includegraphics[width=3.5cm, keepaspectratio]{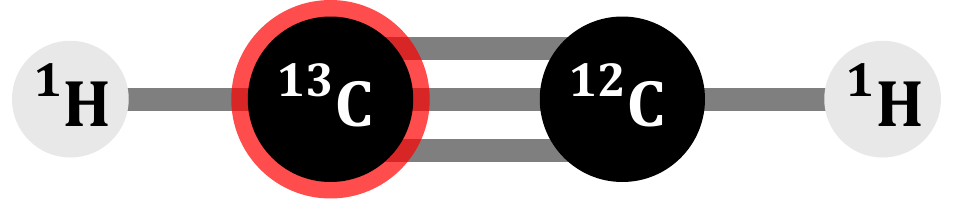}
\end{minipage}
}
  & 1.0000 & 773.484 &\multicolumn{1}{c|}{ \multirow{2}{*}{1.1485} } &  \multirow{2}{*}{728.27}     & \multirow{2}{*}{1.1485} & \multirow{2}{*}{(3)} \\ 
& & 0.9415 & 728.252 & \multicolumn{1}{c|}{}                         &                              &                         &                                            \\  \hline
\multirow{2}{*}{2} &
\multirow{2}{*}{
\begin{minipage}{.3\textwidth}
\centering
\includegraphics[width=3.5cm, keepaspectratio]{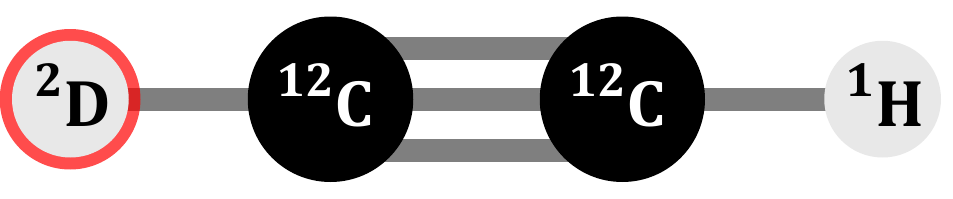}
\end{minipage}
}
  & 1.0000 & 697.315 &\multicolumn{1}{c|}{ \multirow{2}{*}{0.9901} } & \multirow{2}{*}{676.09}       & \multirow{2}{*}{0.9915} & \multirow{2}{*}{(4), (5)}\\
& & 0.9721 & 677.856 &\multicolumn{1}{c|}{}                          &                               &                         &                            \\ \hline
\enddata
\tablerefs{
(1) \citealt{el1999vibrational};
(2) \citealt{herman2007acetylene};
(3) \citealt{fayt2007vibration};
(4) \citealt{herman2004vibration};
(5) \citealt{wlodarczak1989rotational}
}
\end{deluxetable*}

\begin{figure*}[ht!]
  \centering
  \includegraphics[width=0.45\textwidth]{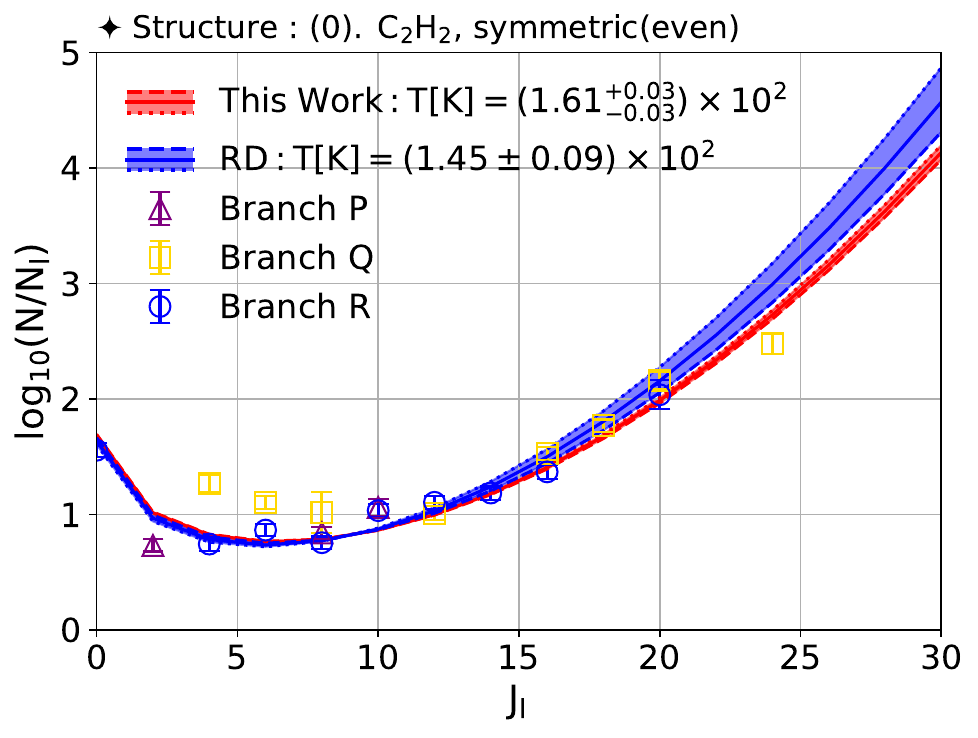}
  \includegraphics[width=0.45\textwidth]{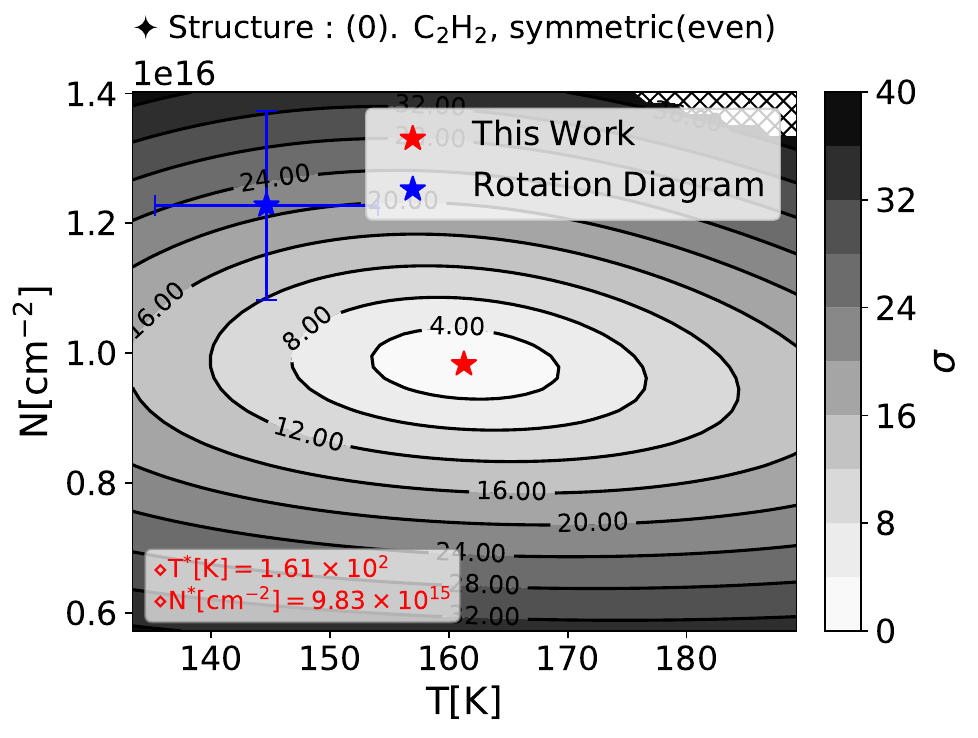}
  \includegraphics[width=0.45\textwidth]{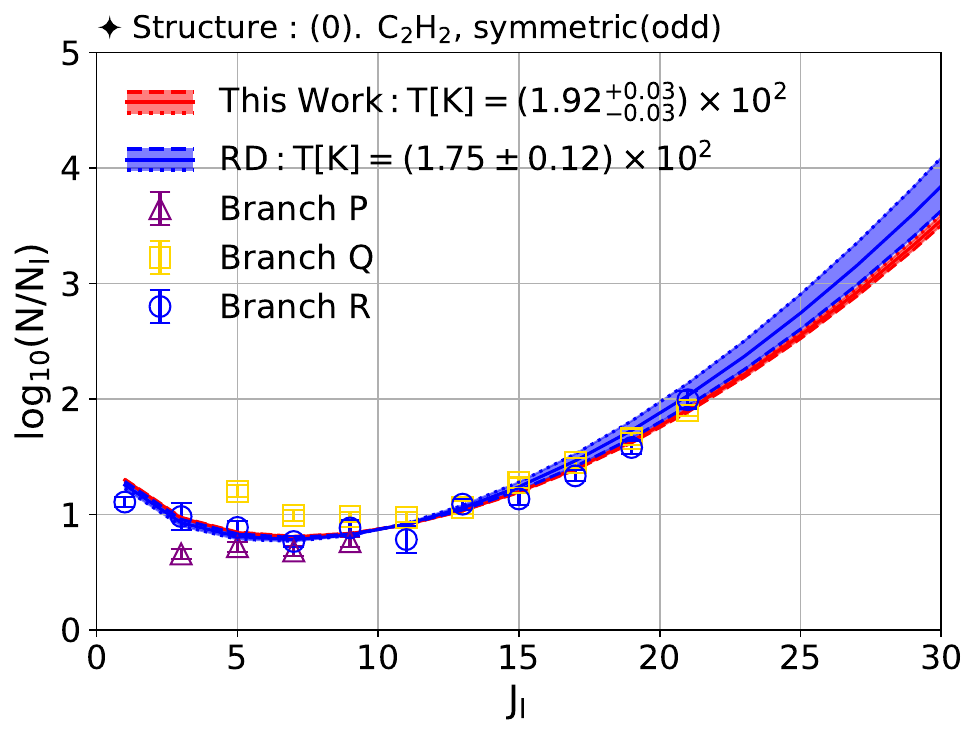}
  \includegraphics[width=0.45\textwidth]{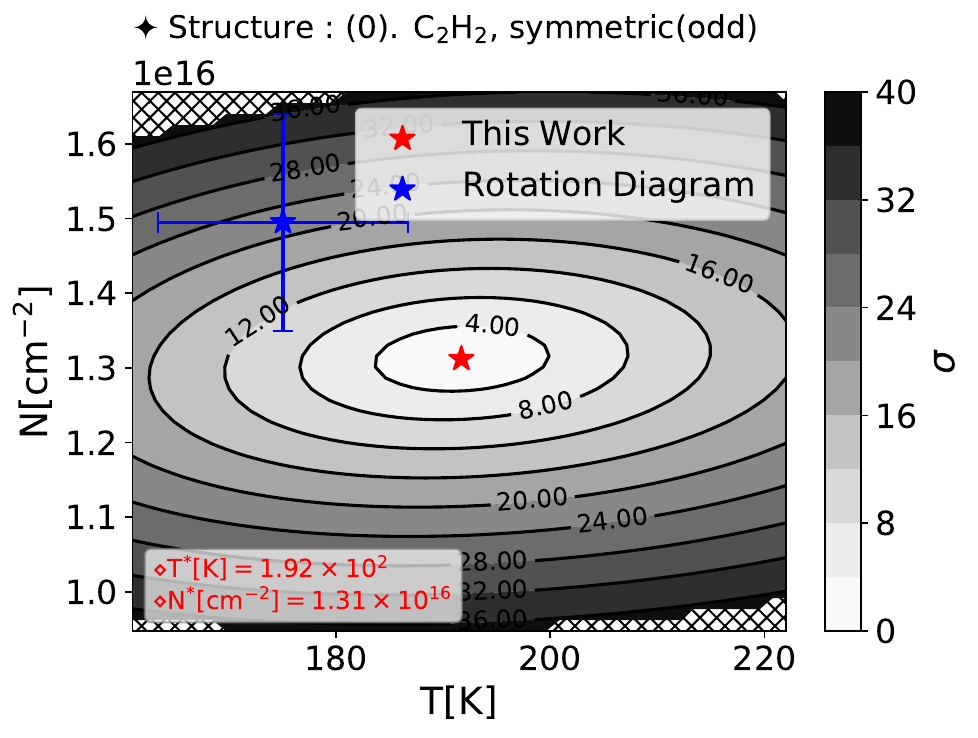}
  \includegraphics[width=0.45\textwidth]{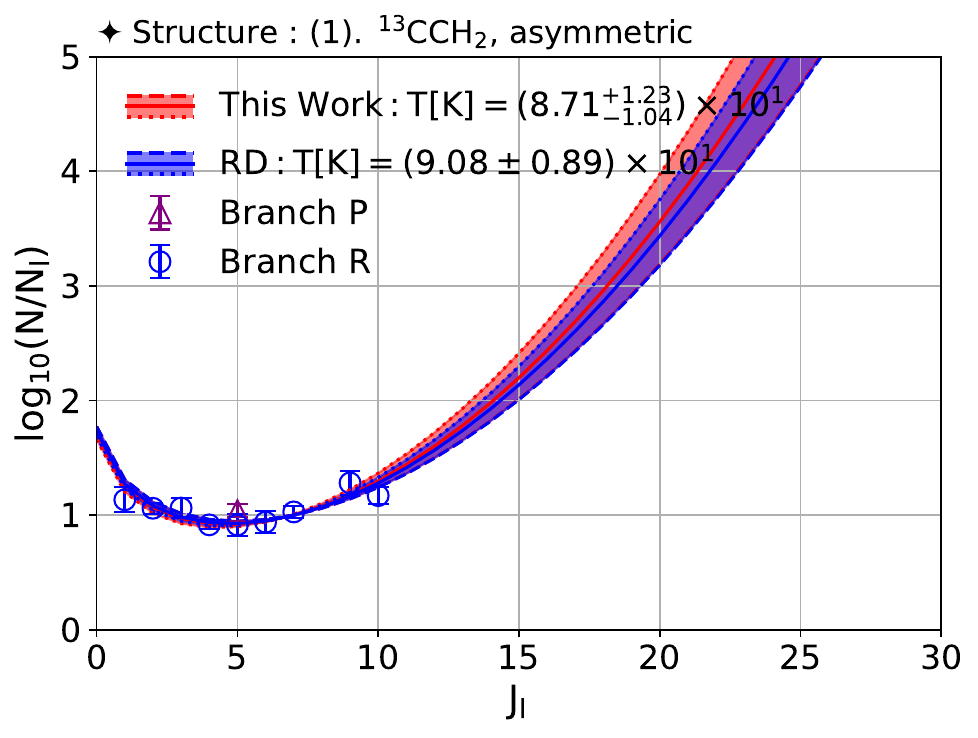}
  \includegraphics[width=0.45\textwidth]{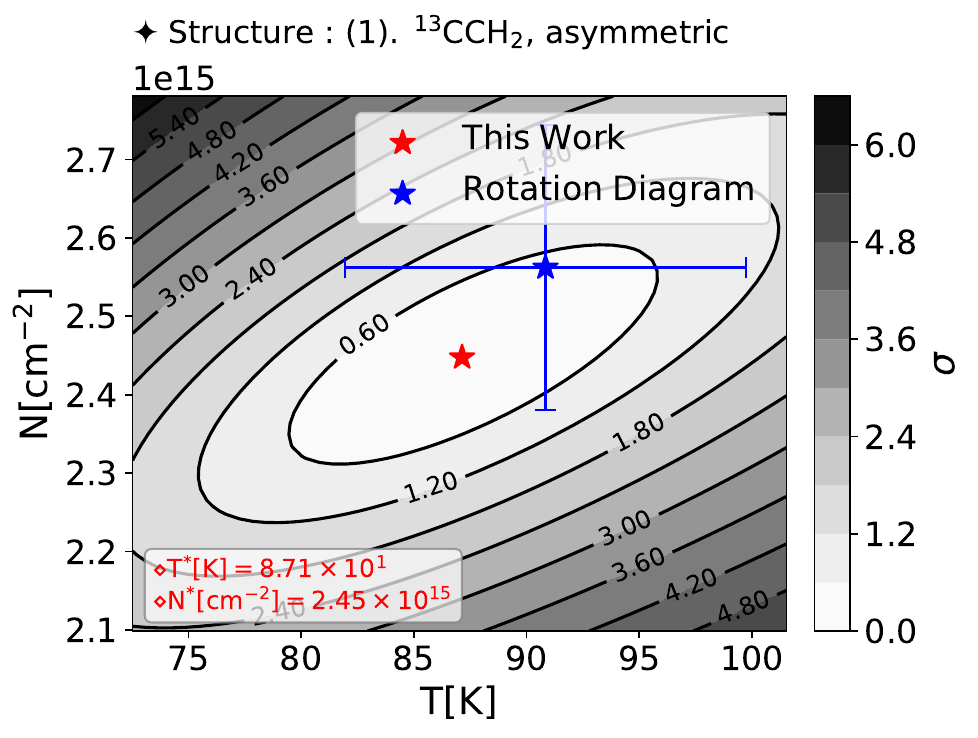}
\caption{
Analysis of the $\nu_5$ vibrational band of acetylene (C$_2$H$_2$) and its isotopologue $^{13}$C$^{12}$CH$_2$ using two approaches: the traditional rotation diagram (RD) method and the $\chi^2$ fitting method. The left panels plot $\log_{10}(N_l/g_l)$ as a function of the lower rotational quantum number $J_l$, while the right panels present contour plots from the $\chi^2$ method.
In the left panels, purple triangles, yellow rectangles, and blue circles indicate observational results for the P-, Q-, and R-branches, respectively, based on the \citet{nickerson2023mid} study of Orion IRc2. Each point includes 1$\sigma$ error bars. The blue shaded regions represent the 1$\sigma$ uncertainty range for the RD method, with the blue solid line showing the best-fit result. The red line represents results from the $\chi^2$ method, providing more precise estimates of physical parameters.
The right panels display contour plots illustrating the confidence intervals for the $\chi^2$ method with two degrees of freedom, temperature (K) and column density (cm$^{-2}$). 
The contours depict the difference between observed and theoretical lower-state column densities ($N_l$), with the minimum value indicating the best-fit parameters. The red star marks the best-fit temperature and column density from the $\chi^2$ method, while blue stars with error bars denote the RD best-fit values with 1$\sigma$ uncertainties. Contour levels correspond to confidence levels, and the explicit best-fit values of $T$ and $N$ are displayed in red in the lower-left corner. These results emphasize the improved accuracy of the $\chi^2$ fitting method (right panels) compared to the RD approach (left panels) in determining the physical properties of the blue clump component.
Additionally, the X-shaped hatching in the right panels indicates regions where 
$\Delta \chi^2$ is too large, making numerical calculations infeasible.
} \label{fig:comp_RD_and_chi2_blue_clump}
\end{figure*}
\begin{figure*}
  \centering
  \includegraphics[width=0.45\textwidth]{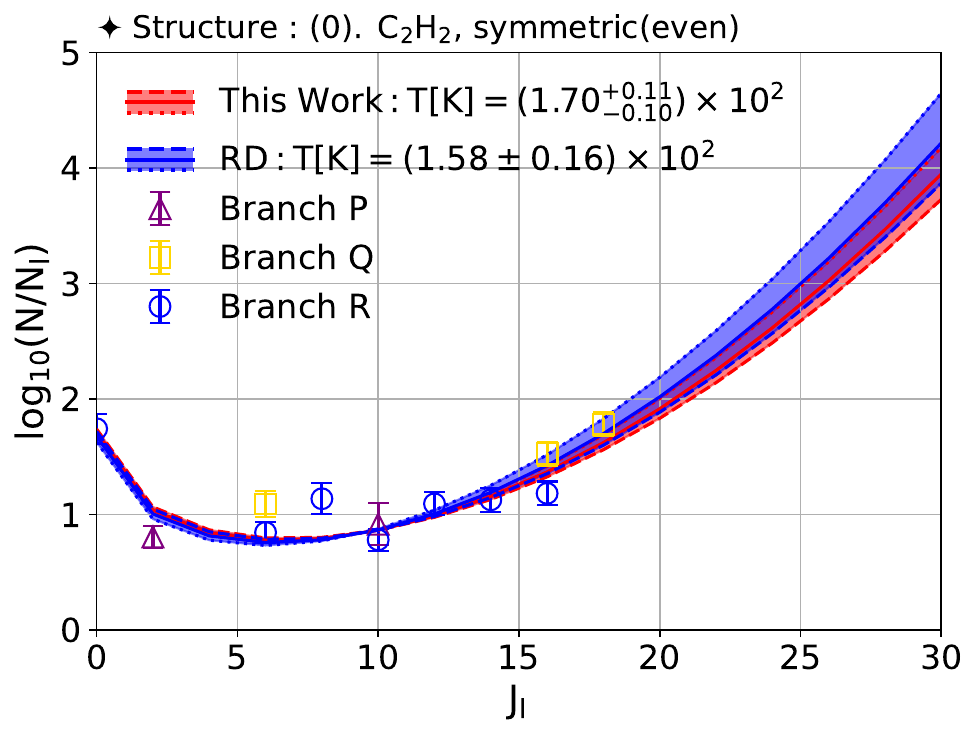}
  \includegraphics[width=0.45\textwidth]{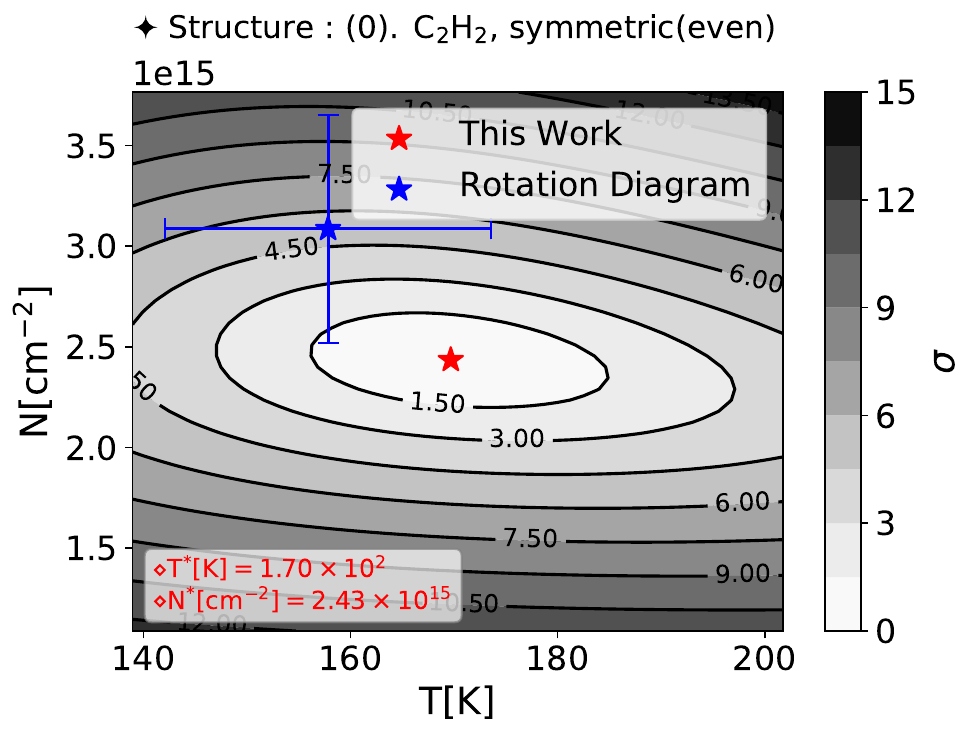}
  \includegraphics[width=0.45\textwidth]{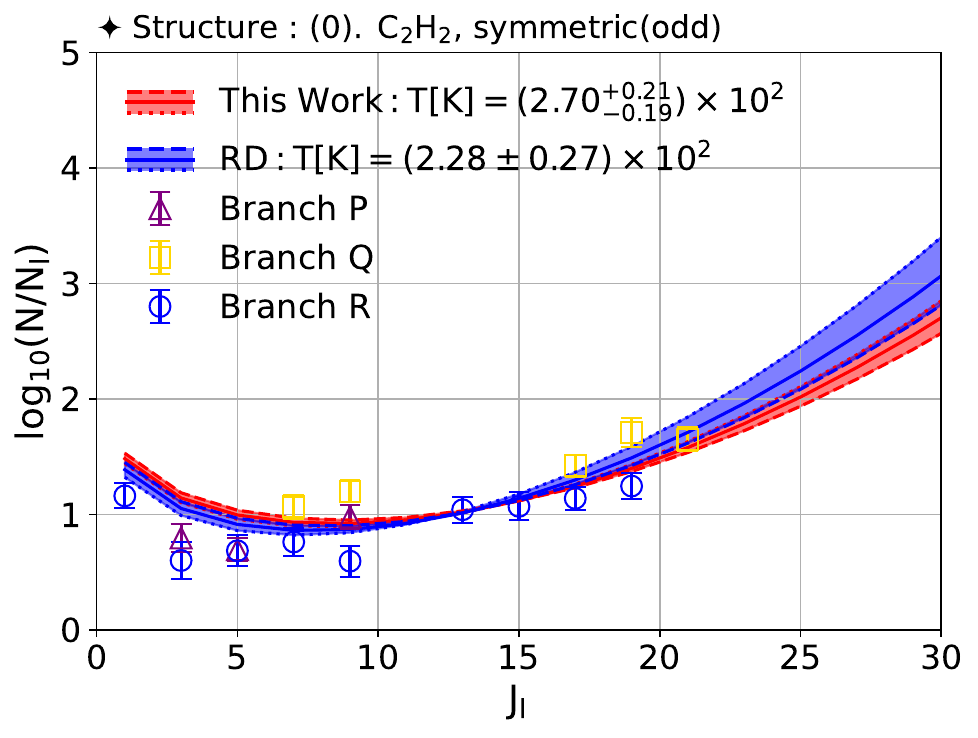}
  \includegraphics[width=0.45\textwidth]{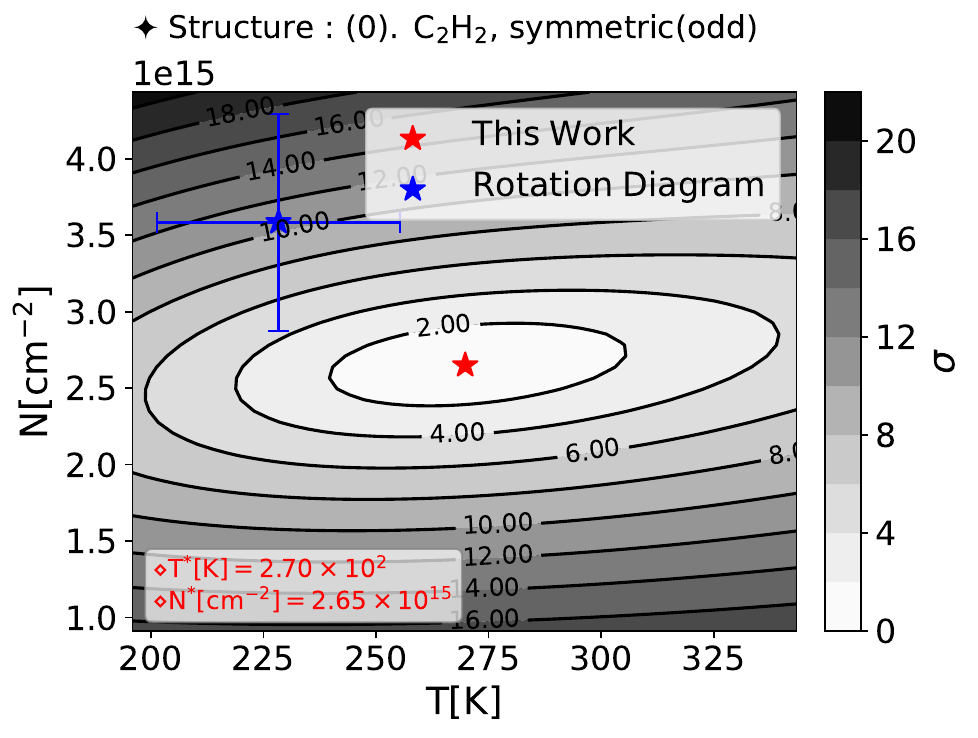}
  \includegraphics[width=0.45\textwidth]{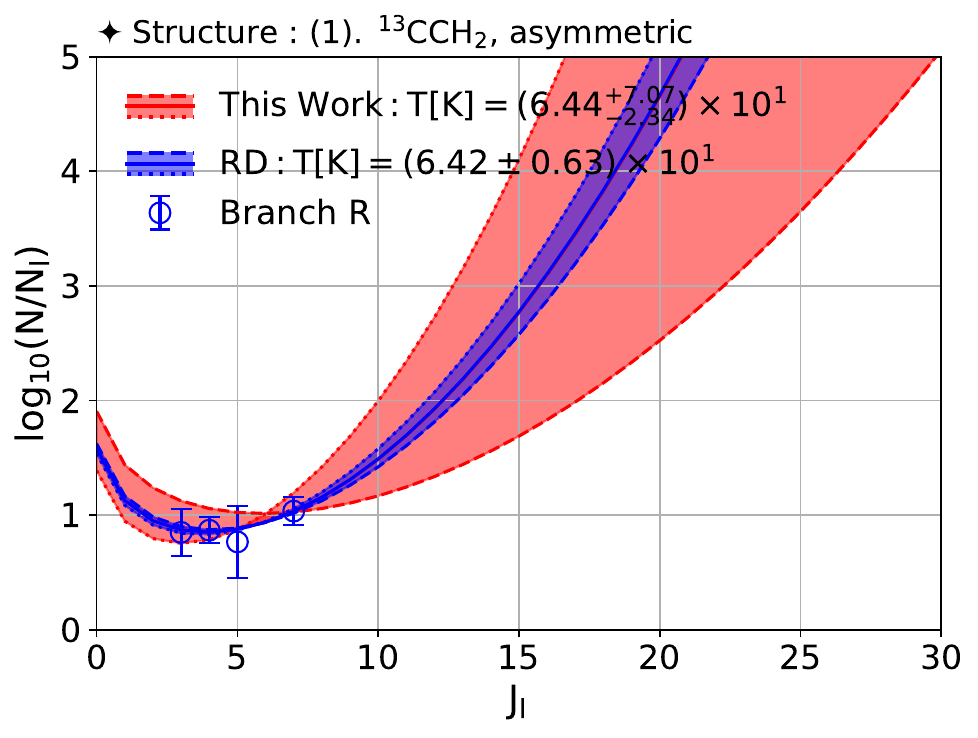}
  \includegraphics[width=0.45\textwidth]{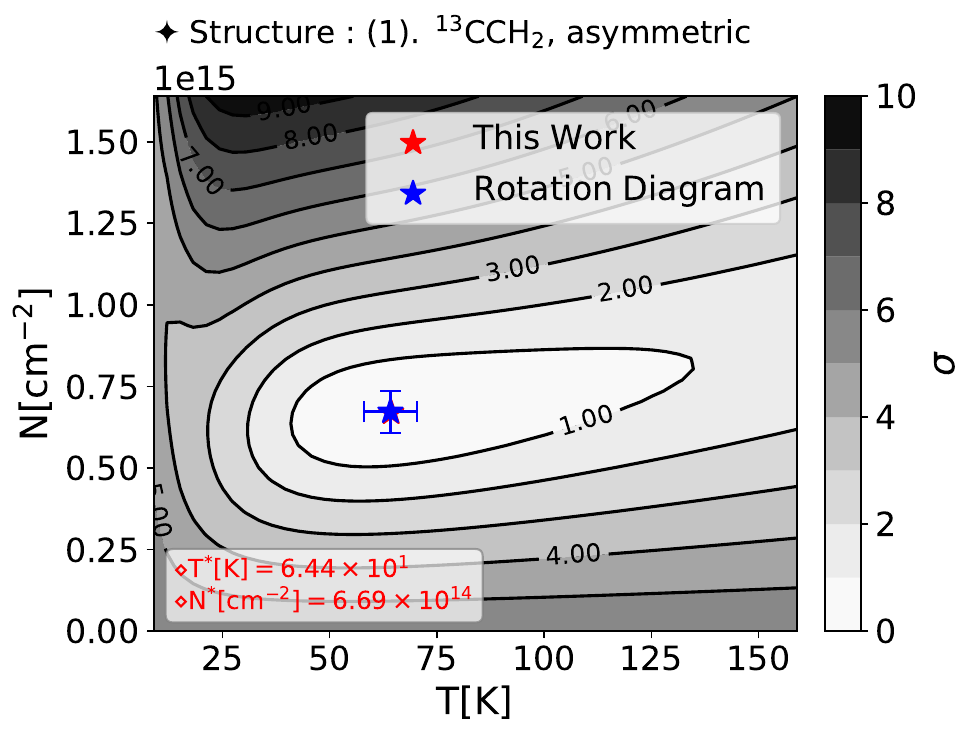}
  \caption{
Same as Figure \ref{fig:comp_RD_and_chi2_blue_clump}, but for the red clump.
} \label{fig:comp_RD_and_chi2_red_clump}
\end{figure*}

\begin{figure}[t!]
  \centering
  \includegraphics[width=0.5\textwidth]{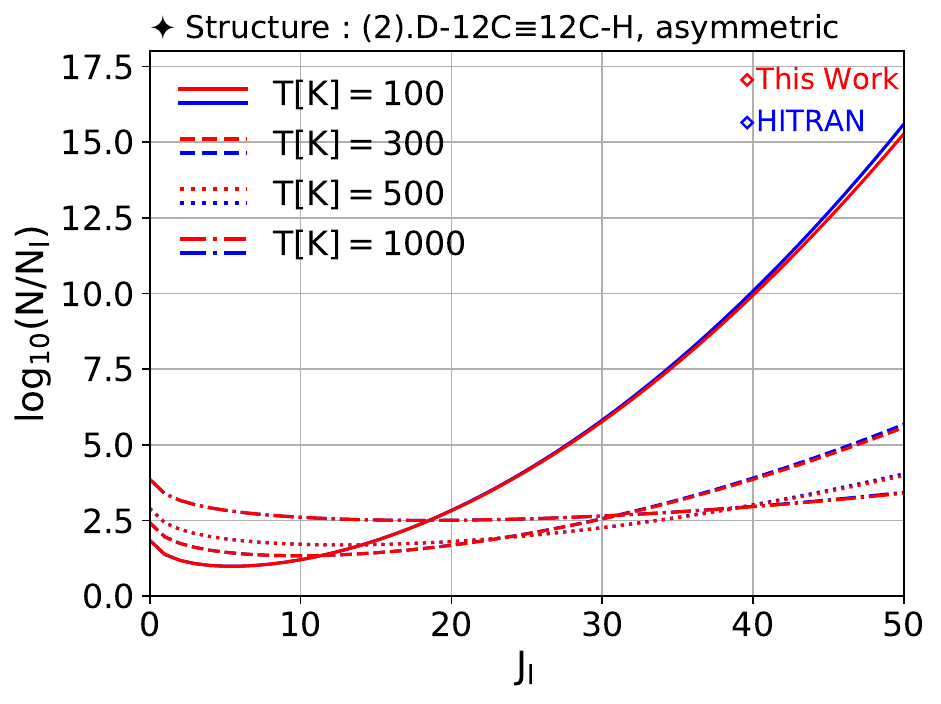}
\caption{
Comparison of calculated $\log_{10}(N_l/g_l)$ values for C$_2$HD as a function of the lower rotational quantum number $J_l$ at different temperatures ($T = 100$ (solid), $300$ (dashed), $500$ (dotted), and $1000$ (dot-dashed) K). The red lines represent the values using our calculated total internal partition (TIP) function, while the blue lines correspond to the reference results using the HITRAN TIP function. The agreement between this work and HITRAN validates the accuracy of the current method, particularly at higher temperatures, where deviations are minimal.
}
\label{fig:log_N_over_Nl_on_C2HD}
\end{figure}

\section{Results}\label{sec:results}
\begin{figure*}[t!]
  \centering
  \includegraphics[width=0.28\textwidth]{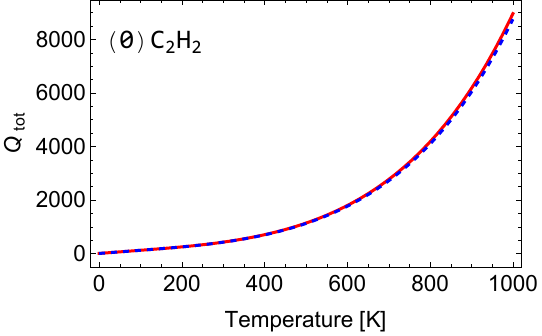}
  \includegraphics[width=0.29\textwidth]{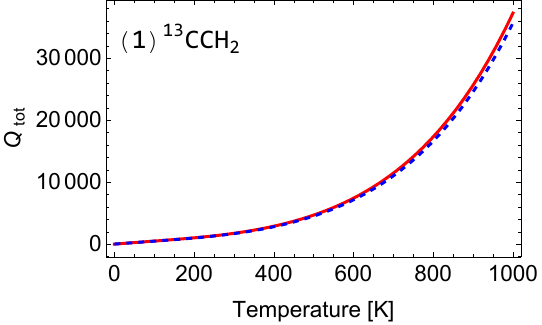}
  \includegraphics[width=0.40\textwidth]{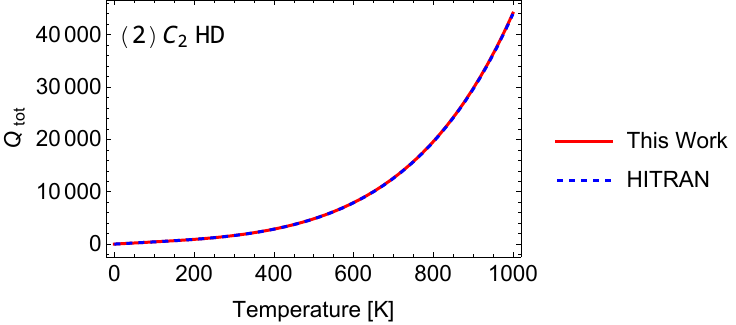}
  \caption{
  Comparison of the total internal partition sums ($Q_\text{tot}$) as a function of temperature for acetylene (C$_2$H$_2$), $^{13}$CCH$_2$, and C$_2$HD. The solid blue line represents the results from this work, while the red dashed line shows the HITRAN values. The agreement between our calculations and HITRAN demonstrates the accuracy of our method across a wide temperature range (0–1000 K). 
  }
\label{fig:ComparisonTIPS}
\end{figure*}
This section presents the application of the $\chi^2$ fitting method for deriving temperature ($T$) and column density ($N$) from observational data. By simultaneously fitting $N$ and $T$, the method provides consistent results for the ortho and para species of C$_2$H$_2$, as well as $^{13}$CCH$_2$. Additionally, theoretical predictions are made for C$_2$HD and other isotopologues not included in the HITRAN database, offering a framework for future studies on these species.

\subsection{$\mathrm{C_{2}H_{2}}$, $\mathrm{^{13}CCH_{2}}$ and $\mathrm{C_{2}HD}$}\label{subsec:results_isotopologues}
\subsubsection{IR spectroscopy}
The HITRAN database provides detailed spectral line parameters for C$_2$H$_2$ and its isotopologues, $^{13}$CCH$_2$ and C$_2$HD, including transition frequencies, line intensities, and broadening coefficients. In the P, Q, and R branches of these species, each transition corresponds to a specific ro-vibrational energy change, which can be directly compared with experimental data or used for modeling. For $^{13}$CCH$_2$, the substitution of $^{12}$C with $^{13}$C results in small spectral line shifts due to the slight increase in mass affecting the rotational constants. In C$_2$HD, the substitution of hydrogen (H) with deuterium (D) causes significantly larger shifts, reflecting the substantial mass difference between hydrogen and deuterium. These isotopic shifts provide valuable information for distinguishing isotopologues in high-resolution spectroscopic analyses. These precise line positions and intensities are crucial for high-resolution spectroscopy, enabling accurate analyses of molecular clouds and planetary atmospheres in astrophysical studies. Instruments like SOFIA/EXES rely on such data for direct comparisons with observed spectra, facilitating detailed investigations of molecular environments.

Table ~\ref{tab:comparison_work} presents the $\nu_5$ vibrational mode frequencies and the rotational constants ($\tilde{B}$) for $\mathrm{C_{2} H_{2}}$ and its isotopologues, $^{13}$CCH$_2$ and C$_2$HD, along with the Fitting Scaling Factor (FSF) values derived from HITRAN data. The FSF values demonstrate consistent scaling across all isotopologues, ensuring the reliability of the vibrational and rotational parameter predictions.
For $^{12}$C$_2$H$_2$, serving as the reference molecule, our calculated and rescaled  $\nu_5$ frequency of 729.185 cm$^{-1}$ (with FSF=0.9413) and rotational constant $\tilde{B}$ of 1.1766 cm$^{-1}$ are in excellent agreement with previously published results (729.08 cm$^{-1}$ and 1.1766 cm$^{-1}$) \citep{el1999vibrational, herman2007acetylene}.
Similarly, for $^{13}$CCH$_2$, the $\nu_5$ frequency of 728.252 cm$^{-1}$ (with FSF=0.9415) and $\tilde{B}$ value of 1.1485 cm$^{-1}$ are also in excellent agreement with the findings of \citet{fayt2007vibration} (728.27 cm$^{-1}$ and 1.1485 cm$^{-1}$). In the case of C$_2$HD, our calculations yield a $\nu_5$ frequency of 677.856 cm$^{-1}$ with FSF=0.9721, 
and a $\tilde{B}$ value of 0.9901 cm$^{-1}$, which are highly consistent with the results reported by \citet{herman2004vibration} and \citet{wlodarczak1989rotational} (676.09 cm$^{-1}$ and 0.9915 cm$^{-1}$).

In this study, the B3LYP DFT method was employed for all quantum chemical calculations \citep{PhysRevA.38.3098, PhysRevB.37.785}. This choice was made after comparing various quantum chemical methods, including HF, B3LYP, PBE0, and MP2, where B3LYP demonstrated the highest accuracy in reproducing known experimental spectra. Consequently, unless otherwise stated, all quantum chemical results presented in this work, including vibrational and rotational parameters, were calculated using the B3LYP method. For further details on the performance and results of other quantum chemical methods, please refer to Appendix~\ref{apd:QCMs} and Appendix~\ref{apd:IR_spectra}.

\subsubsection{Extraction of $T$ and $N$ from $\chi^2$ method}
Under local temperature equilibrium (LTE), the population of lower energy states ($N_l$) is expressed as \citep{goldsmith1999population}:
\[
N_l = N \frac{g_l}{Q} \exp\left(-\frac{E_l}{k_B T}\right),
\]
where $Q$ is the partition function (or TIPS) \citep{GAMACHE1990205, FISCHER2003401, Amway2011Fayt,LARAIA2011391}, and $E_l$ and $g_l$ represent the energy and degeneracy of the lower state, respectively. By taking the natural logarithm, this expression can be transformed into a linear equation:
\[
\ln\left(\frac{N_l}{g_l}\right) = -\frac{1}{T} \left(\frac{E_l}{k_B}\right) + \ln\left(\frac{N}{Q(T)}\right),
\]
which forms the basis of the traditional RD method. In this method, $\ln(N_l / g_l)$ and $E_l / k_B$ are used as $y$ and $x$ coordinates, respectively, to perform a linear fit. While this approach allows for the estimation of temperature ($T$) from the slope and column density ($N$) from the intercept, it assumes that $T$ and $N$ are independent, neglecting the dependence of the partition function $Q$ on $T$.

To overcome this limitation, the $\chi^2$ method is introduced, treating $T$ and $N$ as two variables to be optimized simultaneously. The theoretical values $N_l^{\text{theo}}$ are computed based on $T$ and $N$, and the differences from the observed values $N_l^{\text{obs}}$ are minimized using:
\[
\chi^2 = \sum_{\text{obs } l} \left(\frac{N_l^{\text{theo}} - N_l^{\text{obs}}}{\sigma_{\text{obs}}}\right)^2,
\]
where $\sigma_{\text{obs}}$ accounts for observational uncertainties. By finding the $T$ and $N$ values that minimize $\chi^2$, the method provides a more robust and accurate estimation of physical parameters.

The analysis of acetylene (C$_2$H$_2$) and its isotopologue $^{13}$C$^{12}$CH$_2$ was conducted for two distinct velocity components, referred to as the blue clump and red clump. These components are characterized by their unique velocity shifts and allow for the assessment of the physical conditions of the molecular gas. In this section, we focus on the blue clump, which is distinguished by its relatively narrow velocity dispersion and prominent ro-vibrational features.

Figure~\ref{fig:comp_RD_and_chi2_blue_clump} presents the analysis of the blue clump
component for acetylene (C2H2) and its isotopologue $^{13}$CCH$_{2}$ using two methods: the traditional RD method and the newly implemented $\chi^2$ fitting method for the blue clump component. The left panels display $\log_{10}(N_l/g_l)$ as a function of the lower rotational quantum number $J_l$, derived from SOFIA high-resolution observations, with data from \citet{nickerson2023mid}. Transitions in the P-, Q-, and R-branches are represented by purple triangles, yellow rectangles, and blue circles, respectively, with 1$\sigma$ error bars included for each point. The blue shaded regions indicate the uncertainty range associated with the RD method, while the blue solid line represents the best-fit result obtained through linear fitting. In contrast, the red line shows the best-fit result derived from the $\chi^2$ method, demonstrating greater accuracy and reliability.

The right panels illustrate the confidence intervals for the $\chi^2$ method with two degrees of freedom, temperature ($T$) and column density ($N$).
These plots visualize the agreement between the observed and theoretical lower-state column densities ($N_l$). The minimum contour value marks the best-fit parameters, with red stars indicating the results from the $\chi^2$ method and blue stars representing those from the RD method. The associated error bars for the RD method are also shown. The best-fit values of $T$ and $N$ are explicitly stated in red text in the lower-left corner of each panel.
This analysis demonstrates the effectiveness of the $\chi^2$ method in interpreting complex molecular spectra. Compared to the RD approach, the $\chi^2$ method achieves a significantly lower $\chi^2$ value, highlighting its superior reliability in determining the physical conditions within the blue clump.

Figure \ref{fig:comp_RD_and_chi2_red_clump} extends the analysis to the red clump component for acetylene (C$_2$H$_2$) and its isotopologue $^{13}$CCH$_2$, following the same methodology as used for the blue clump. The red clump represents another distinct velocity component, enabling a complementary investigation of its physical properties.

The analysis of the red and blue clumps reveals distinct differences in parameter uncertainties and fitting performance between the $\chi^2$ and RD methods. For the blue clump, the $\chi^2$ method demonstrates a significant improvement by reducing error bars by up to
75\%
compared to the RD method. This reduction highlights the robustness and precision of the $\chi^2$ approach, which simultaneously fits temperature ($T$) and column density ($N$) while accounting for statistical weights and degeneracy factors. The improved precision underscores the advantage of global fitting over linear extrapolation in traditional RD methods, particularly for symmetric molecules like C$_2$H$_2$.
In contrast, the red clump shows larger uncertainties for $^{13}$CCH$_2$ when using the $\chi^2$ method compared to the RD method. This is evident in the left panel of Figure \ref{fig:comp_RD_and_chi2_red_clump}, where the red band (representing the uncertainty range from the $\chi^2$ method) is noticeably broader than the blue band (uncertainty range from the RD method). The increase in uncertainty arises from the limited number of observational data points (four) used in the fitting process. In the $\chi^2$ fitting method, the uncertainty in the parameters is inversely proportional to the square root of the degrees of freedom, calculated as $N_{\text{data}} - N_{\text{parameters}}$. With only four data points and two fitting parameters ($T$ and $N$), the degrees of freedom are minimal ($2$), leading to inherently larger uncertainties in the $\chi^2$ fitting compared to cases with higher observational coverage. This highlights the critical role of sufficient data points to achieve reduced uncertainties in multi-parameter fitting methods.

Despite the larger uncertainties observed in the $\chi^2$ method for the red clump, it is notable that the center values of $^{13}$CCH$_2$ derived from both the RD and $\chi^2$ methods are remarkably similar. 

These findings underscore the complementary nature of the RD and $\chi^2$ methods. While the RD method provides a quick estimation with smaller uncertainties in cases of sparse data, the $\chi^2$ method offers a more detailed and statistically consistent analysis, making it particularly valuable for datasets with higher observational coverage. Moving forward, improving data quality and quantity will be crucial to fully leverage the precision and reliability of advanced fitting methods like $\chi^2$ in astrophysical studies.

Figure \ref{fig:log_N_over_Nl_on_C2HD} compares the calculated $\log_{10}(N_l/g_l)$ values for C$_2$HD as a function of the lower rotational quantum number $J_l$ at four temperatures ($T = 100$ (solid), $300$ (dashed), $500$ (dotted), and $1000$ (dot-dashed) K). The red lines represent the results using the TIP function calculated in this work, while the blue lines correspond to reference values based on the HITRAN TIP function.

The results show excellent agreement across the full range of $J_l$, especially at higher temperatures ($T = 500$ K and $1000$ K), where deviations are negligible. Slight discrepancies at lower temperatures ($T = 100$ K and $300$ K) for higher $J_l$ values likely arise from differences in partition function calculations. However, these deviations remain well within acceptable limits, confirming the accuracy and robustness of the TIP function derived in this study. This comparison highlights the reliability of the proposed approach for spectroscopic studies and its consistency with established databases like HITRAN.
For a detailed discussion on the TIP function and its calculations, refer to Section ~\ref{subsubsec:TIPS}.


\subsection{The remaining isotopologues of acetylene not mentioned in HITRAN}\label{subsec:results_theoretical}
HITRAN is widely trusted in the field due to its comprehensive and accurate data, which is crucial for spectroscopic analyses. However, for acetylene, HITRAN provides only three types of TIPS (Total Internal Partition Sums) \citep{GAMACHE2021107713}, which limits the data available for isotopologue studies. To overcome this, we performed additional quantum chemical calculations using \texttt{Gaussian16} for a total of seven acetylene isotopes (excluding the two in the previous section),
$^{13}$C$_2$H$_2$, H$^{13}$C$^{12}$CD, H$^{12}$C$^{13}$CD, $^{12}$C$_2$D$_2$, $^{13}$C$_2$HD, $^{12}$C$^{13}$CD$_2$, and $^{13}$C$_2$D$_2$)
using four different quantum chemistry methods: HF, DFT with B3LYP and PBE0, and MP2.
For quantum chemical calculation methodology, see Appendix \ref{apd:QCMs}; 
for fundamental frequencies corresponding to 
$\nu_1 \sim \nu_5$ and detailed calculation results and discussion, see Appendix \ref{apd:IR_spectra}.
\subsubsection{Total Internal Partition Sum (TIPS)}\label{subsubsec:TIPS}
The TIPS of a molecule represents the sum of contributions from all possible energy states and is crucial for accurately modeling molecular properties across a range of temperatures. It can be calculated as:
\begin{equation}
	Q(T) = \sum_i \exp \left( \frac{-\epsilon_i}{k_B T} \right)~.
\end{equation}
where \( \epsilon \) represents the energy levels. These levels are typically decomposed into rotational and vibrational components
\footnote{More precisely, $\epsilon = \epsilon_{trans} + \epsilon_{rot} + \epsilon_{vib} + \epsilon_{elec} + \epsilon_{others}$. The $\epsilon_{trans(elec)}$ is corresponding to 
the translational(electronical) energy levels. 
Note that the term $\epsilon_{others}$ may represent the nuclear spin energy (levels) 
and/or the interactions between the first fours.
We here assume that the first fours are independent, and the last terms are negligible.
However, note that the energy levels can not be independent and can become coupled to each other as the energy of the molecule increases.
Since \(\epsilon_{elec} > \epsilon_{vib} > \epsilon_{rot} \gg \epsilon_{trans}\), translational states far outnumber rotational, vibrational, and electronic states. Typically, \(q_{elec}\) is close to unity, while \(q_{vib}\) and \(q_{rot}\) range from 1 to 100, and \(q_{trans}\) can exceed \(10^{20}\).
}:
\begin{equation}
	\epsilon = \epsilon_{rot} + \epsilon_{vib}~. \label{eq:Energy_Molecule}
\end{equation}
Using this separated energy levels, 
the total partition function is expressed as:
\begin{equation}
	Q_{total} = Q_{rot} \cdot Q_{vib} ~.
\end{equation}
For linear molecules such as acetylene, 
the rotational energy form can be expressed as follows:
\begin{eqnarray}
    \epsilon_J = \frac{\hbar^2 J(J+1)}{2I} &=& h c \,\tilde{B} J(J+1) - \tilde{D}_J  J^2(J+1)^2 ~, \nonumber \\
    \mathrm{where}~ J &=& 0, 1, 2, 3, 4, \cdots 
\end{eqnarray}
where $\tilde{B}$ is the rotation constant expressed in $\rm{cm}^{-1}$, 
\begin{equation}
	\tilde{B} = \frac{\hbar^2}{2 I hc} = \frac{h}{8 \pi^2 I c}~,
\end{equation}
and \( \tilde{D}_J \) is the centrifugal distortion constant. 
The rotational constants ($\tilde{B}$) for each isotopologue of acetylene were determined using quantum chemical calculations based on the B3LYP method. For $^{12}$C$_2$H$_2$, $^{12}$C$^{13}$CH$_2$, and $^{12}$C$_2$HD, the FSF values for $\nu_5$ and $\tilde{D}_J$ were optimized to match the observed spectra from HITRAN. For the remaining isotopologues, FSFs were estimated based on structural similarities. Table~\ref{tab:used_constants} summarizes the FSF, $\tilde{B}$, and $\tilde{D}_J$ values used for acetylene and its isotopologues in this study.

\begin{deluxetable}{cccc}[h!]
\tablecaption{Rotational constants ($\tilde{B}$) and centrifugal distortion constants ($\tilde{D}_J$) for acetylene ($^{12}$C$_2$H$_2$) and its isotopologues, determined using quantum chemical calculations based on the B3LYP method. FSF values were adjusted to best fit the observed $\nu_5$ band spectra in HITRAN for $^{12}$C$_2$H$_2$, $^{12}$C$^{13}$CH$_2$, and $^{12}$C$_2$HD, while FSFs for other isotopologues were estimated based on structural similarities. The $\tilde{B}$ values are expressed in cm$^{-1}$, and $\tilde{D}_J$ values are scaled by $10^{-6}$ for clarity.}
\label{tab:used_constants}
\tablehead{
    \colhead{\textbf{Molecule ID}} & \colhead{\textbf{FSF}} & \colhead{$\tilde{B}$ (cm$^{-1}$)} & \colhead{$\tilde{D}_J$ ($\times 10^{-6}$ cm$^{-1}$)}     
}
\startdata
0 & 0.9413 & 1.1766 & 12.2538 \\
1 & 0.9415 & 1.1485 & 11.4259 \\
2 & 0.9721 & 0.9901 & 8.4493 \\
3 & 0.9413 & 1.1197 & 10.6241 \\
4 & 0.9721 & 0.9659 & 7.9032 \\
5 & 0.9721 & 0.9740 & 8.0506 \\
6 & 0.9413 & 0.8457 & 8.4405 \\
7 & 0.9721 & 0.9491 & 7.5049 \\
8 & 0.9413 & 0.8310 & 8.0511 \\
9 & 0.9413 & 0.8160 & 7.6671 \\
\enddata
\end{deluxetable}

\twocolumngrid
\begin{deluxetable*}{ccccccccc}
\tablecaption{Summary table for each isotopical species, including molecular ID, isotopical name, nuclear spin values, total degeneracy, symmetry statistics, rovibronic and nuclear spin wavefunctions, $J$ parity, and statistical weights ($g_i$). For detailed explanations, including symmetry constraints and 
their impact on statistical weights, refer to the main text.}
\label{tab:g_i}
\tablehead{
    \colhead{Molecule} & \colhead{Isotopical} & \colhead{$I(C_1 + H_1)$} & \colhead{$I(C_2 + H_2)$} & \colhead{Total (a)} & \colhead{Statistic} & \colhead{$\Psi_{rov} \times \Psi_{ns} = \Psi_{tot}$} & \colhead{$J$} & \colhead{Statistical} \\     
    \colhead{ID} & \colhead{species} & \colhead{} & \colhead{} & \colhead{degeneracy} & \colhead{} & \colhead{} & \colhead{parity} & \colhead{weight ($g_i$)}
}
\startdata
0 & \ce{C2H2}              & 1/2        & 1/2      & 4  & Fermi (a) & $s \times a = a$ & even      & 1 \\
  &                        &            &          &    &           & $a \times s = a$ & odd       & 3 \\ \hline
1 & \ce{H ^{13}C ^{12}C H} & 1,0        & 1/2      & 8  &  \dots    & \dots            & even, odd & 8 \\ \hline
2 & \ce{C2DH}              & 1/2        & 1        & 6  &  \dots    & \dots            & even, odd & 6 \\ \hline
3 & \ce{^{13}C2 H2}        & 1,0        & 1,0      & 16 & Bose (s)  & $s \times s = s$ & even      & 10 \\
  &                        &            &          &    &           & $a \times a = s$ & odd       & 6 \\ \hline
4 & \ce{D ^{12}C ^{13}C H} & 1          & 0, 1     & 12 & \dots     & \dots            & even, odd & 12 \\ \hline
5 & \ce{D ^{13}C ^{12}C H} & 1/2, 3/2   & 1/2      & 12 & \dots     & \dots            & even, odd & 12 \\ \hline
6 & \ce{C2D2}              & 1          & 1        & 9  & Bose (s)  & $s \times s = s$ & even      & 6 \\
  &                        &            &          &    &           & $a \times a = s$ & odd       & 3 \\ \hline
7 & \ce{D ^{13}C ^{13}C H} & 1/2, 3/2   & 0, 1     & 24 &  \dots    & \dots            & even, odd & 24 \\ \hline
8 & \ce{D ^{13}C ^{12}C D} & 1/2, 3/2   & 1        & 18 &  \dots    & \dots            & even, odd & 18 \\ \hline
9 & \ce{^{13}C2 D2}        & 1/2, 3/2   & 1/2, 3/2 & 36 & Fermi (a) & $s \times a = a$ & even      & 15 \\
  &                        &            &          &    &           & $a \times s = a$ & odd       & 21 \\ \hline
\enddata
\end{deluxetable*}

The centrifugal distortion term ($\tilde{D}_J$) plays a more significant role at large values of \( J \), opposing the leading rotational constant term and introducing potential deviations in the calculated energy levels. While higher-order correction terms are necessary to fully account for these effects, they are omitted in this study to prioritize computational efficiency and focus on primary contributions. This simplification may explain discrepancies observed at high \( J \) values, especially for isotopologues with pronounced centrifugal distortion effects.

With this energy spectrum, the rotational partition function is calculated as follows:
\begin{equation}
	Q_{rot} = g_i \sum_{J=1}^{\infty} g_J \exp \left(-\frac{\epsilon_J}{k_B T} \right)~,
\end{equation}
where the $g_J = (2J+1) $ is the degeneracy factor of the given energy $\epsilon_J$. The $g_i$ describe the state-independet degeneracy factor.
For acetylene, we use the ``astronomer convention'' in which para:ortho species ratio is 1/4:3/4, resulting in \( g_i^{astro} = 1 \). In contrast, the ``atmospheric convention'' assumes a para:ortho ratio of 1:3, yielding \( g_i^{atmos} = 4 \). Thus, the \(g_i^{atmos} \) value in the atmospheric convention is 4 times larger than in the astronomer convention ($g_i^{atmos} = 4 \times g_i^{astro}$).
For all other molecules, the atmospheric convention was used. Therefore, when using other conventions, such as the astronomer convention, it is suggested to multiply by an appropriate constant.

The vibrational energy levels are given by
\begin{equation}
	\epsilon_v = h \nu \left( v + \frac{1}{2} \right) = hc \bar{\nu} \left( v + \frac{1}{2} \right)~,~~~ v =0, 1, 2, 3, \cdots
\end{equation}
The $\nu$ represents the vibrational frequency of the molecules,
\begin{equation}
	\nu = \frac{1}{2\pi} \left( \frac{k}{\mu}  \right)^{\frac{1}{2}}~,
\end{equation}
where the $k$ is the spring constant of the molecule and $\mu$ is the reduced mass of the same molecule.
And the $\bar{\nu} \equiv (\nu /c) $ is the wave number in $\rm{cm}^{-1}$.

The vibrational partition function of a linear molecule is
\begin{equation}
	Q_{vib} = \frac{1}{1 - \exp\left(-\frac{h\nu}{k_B T} \right)} =  \frac{1}{1 - \exp\left(-\frac{h c \bar{\nu}}{k_B T} \right)}
\end{equation}

Table \ref{tab:g_i} summarizes the properties of various isotopologues of acetylene, including molecular ID, isotopical name, nuclear spin values for each atom ($I(\mathrm{C_1+H_1})$ and $I(\mathrm{C_2+H_2})$), total degeneracy, symmetry statistics (Fermi or Bose), rovibronic and nuclear spin wavefunctions ($\Psi_{\mathrm{rov}} \times \Psi_{\mathrm{ns}} = \Psi_{\mathrm{tot}}$), $J$ parity (even or odd), and statistical weights ($g_i$). The total degeneracy is calculated using the formula $(2I(\mathrm{C_1+H_1}) + 1)(2I(\mathrm{C_2+H_2}) + 1)$, which accounts for the contributions of nuclear spin states for the relevant nuclei.
For symmetric isotopologues such as C$_2$H$_2$, \ce{^{13}C2 H2}, C$_2$D$_2$ and \ce{^{13}C2 D2} \citep{Herman2003Campargue}, specific symmetry constraints result in distinct statistical weights for even and odd $J$ states, following Fermi-Dirac or Bose-Einstein statistics. 
In contrast, asymmetric isotopologues, such as 
H$^{13}$C$^{12}$CH ($^{13}$CCH$_{2}$), C$_{2}$DH, D$^{12}$C$^{13}$CH, D$^{13}$C$^{12}$CH, 
D$^{13}$C$^{13}$CH ($^{13}$C$_{2}$DH), D$^{13}$C$^{12}$CD ($^{13}$CCD$_{2}$), 
exhibit altered nuclear spin configurations that lead to higher degeneracies and statistical weights due to the larger number of possible spin states. These $g_i$ factors are critical for calculating molecular partition functions, understanding the relative populations of ro-vibrational states, and accurately modeling molecular spectra.

Figure \ref{fig:ComparisonTIPS} presents a comparison of the total internal partition sums ($Q_\text{tot}$) as a function of temperature for acetylene (C$_2$H$_2$), $^{13}$C$^{12}$CH$_2$, and C$_2$HD. The solid blue lines represent the partition sums calculated in this work, while the red dashed lines correspond to the values provided by HITRAN. Across the temperature range of 0 to 1000 K, our results show excellent agreement with the HITRAN data, confirming the reliability of our method for calculating partition sums. The slight deviations at higher temperatures, particularly for $^{13}$C$^{12}$CH$_2$, can be attributed to the different computational approaches used. Overall, these comparisons validate the robustness of our approach in predicting partition sums for acetylene isotopologues, ensuring accurate results for future spectral analysis.
For TIP results for the remaining seven isotopoloues of acetylene, 
see appendix \ref{apd:TIP}.

\begin{figure*}[th!]
  \centering
  \includegraphics[width=0.4\textwidth]{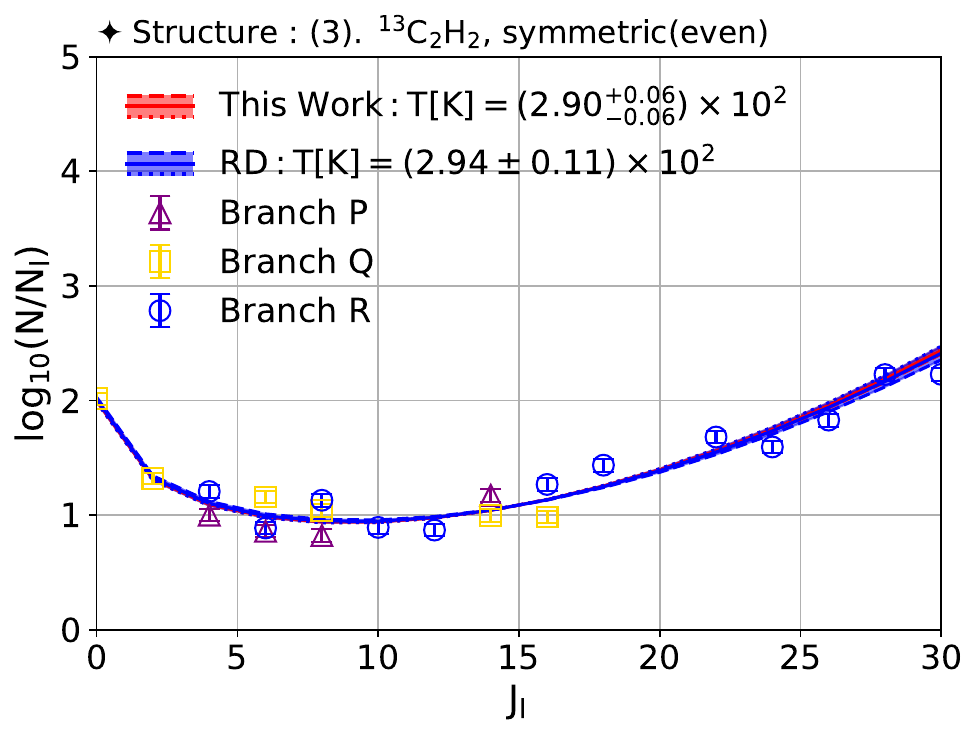}
  \includegraphics[width=0.4\textwidth]{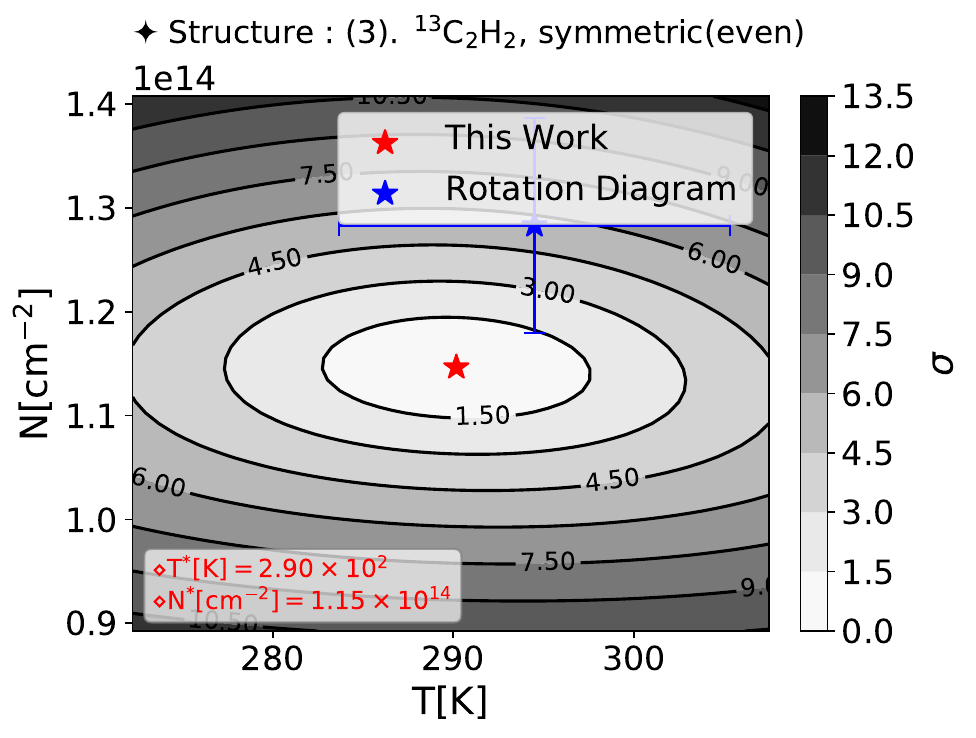}
  \includegraphics[width=0.4\textwidth]{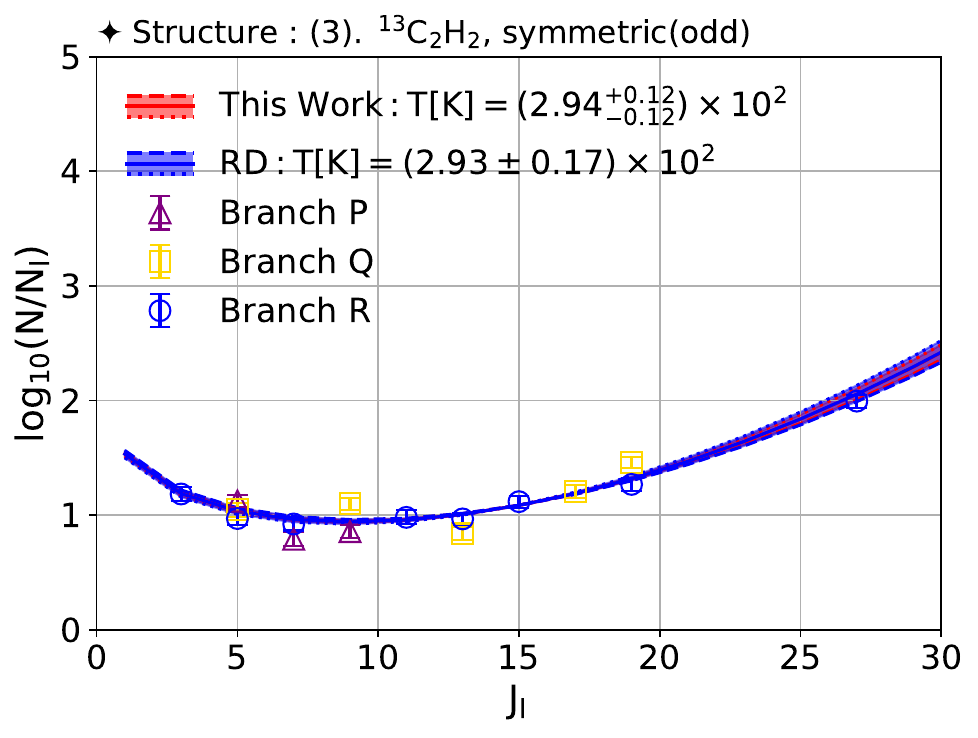}
  \includegraphics[width=0.4\textwidth]{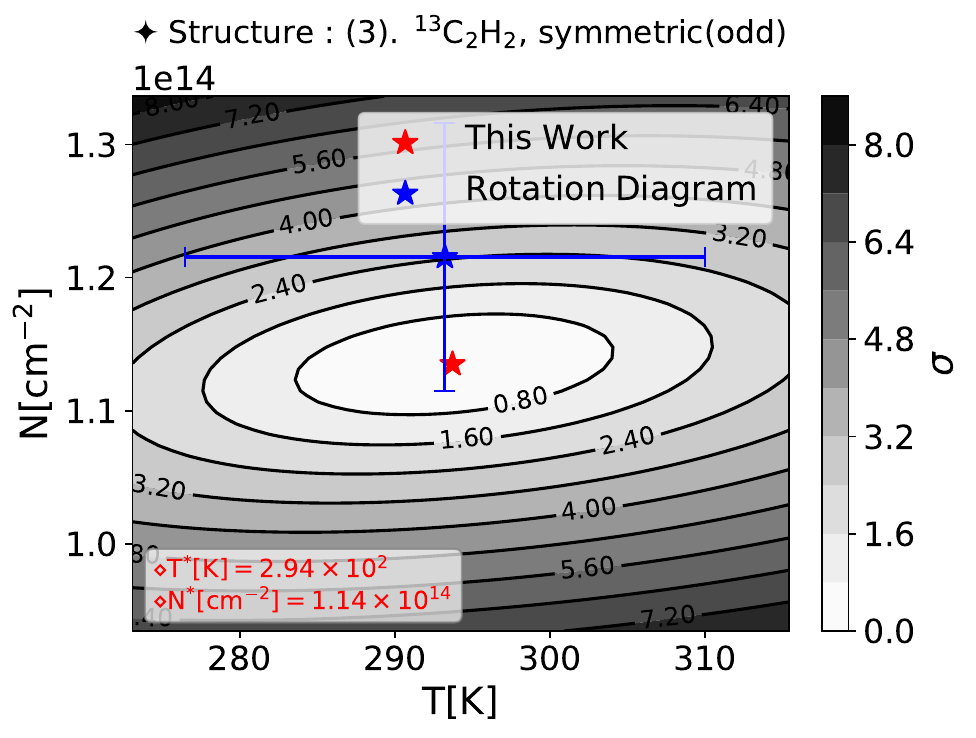}
\caption{
Same as Figures \ref{fig:comp_RD_and_chi2_blue_clump} and \ref{fig:comp_RD_and_chi2_red_clump}, but for $^{13}$C$_2$H$_2$ (molecule ``3''). 
For more details, refer to the main text.
}\label{fig:13C2H2}
\end{figure*}

\subsubsection{$\chi^2$ methods for the remaining acetylene isotoplogues}
Here, we consider $^{13}$C$_2$H$_2$ (molecule ``3'' in our convention) as a specific example to evaluate the performance of our two-parameter fitting method using chi-squared ($\chi^2$) calculations. In the absence of actual observational data, we generated a pseudo dataset to demonstrate how the method can be applied to isotopologues with unique statistical properties. This analysis highlights the versatility of our approach in determining temperature and column density, and provides a framework for studying astrophysical environments such as carbon-rich stars or regions with high $^{13}$C abundance.

Figure \ref{fig:13C2H2} shows the results for $^{13}$C$_2$H$_2$, focusing on the estimation of temperature ($T$) and column density ($N$). The left panels present $\log_{10}(N_l/g_l)$ as a function of the lower rotational quantum number $J_l$, with purple triangles, yellow rectangles, and blue circles representing P-, Q-, and R-branch transitions, respectively. The blue shaded regions indicate the 1$\sigma$ uncertainty from the RD method, while the blue solid line denotes the best-fit result using this method. The red line corresponds to the best-fit result derived from our two-parameter fitting method, demonstrating the improved accuracy and flexibility of the latter.
The right panels illustrate the confidence intervals for the $\chi^2$ method with two degrees of freedom, temperature ($T$) and column density ($N$).
The right panels display contour plots illustrating the confidence intervals for the $\chi^2$ method with two degrees of freedom, $T$ and $N$.
Darker regions represent higher $\Delta \chi^2$ values, while lighter regions correspond to the lower $\Delta \chi^2$, which provides the best estimates for $T$ and $N$. Unlike the C$_2$H$_2$ (molecule ``0''), $^{13}$C$_2$H$_2$ (molecule ``3'') has a flipped statistical weight due to its nuclear spins and Boson statistics, 
where even rotational states have a larger weight than odd states. This is reflected in the distinct separation of fitting results for even and odd rotational states, highlighting the molecule's unique symmetry properties.

\begin{figure*}[t!]
  \centering
  \includegraphics[width=0.35\textwidth]{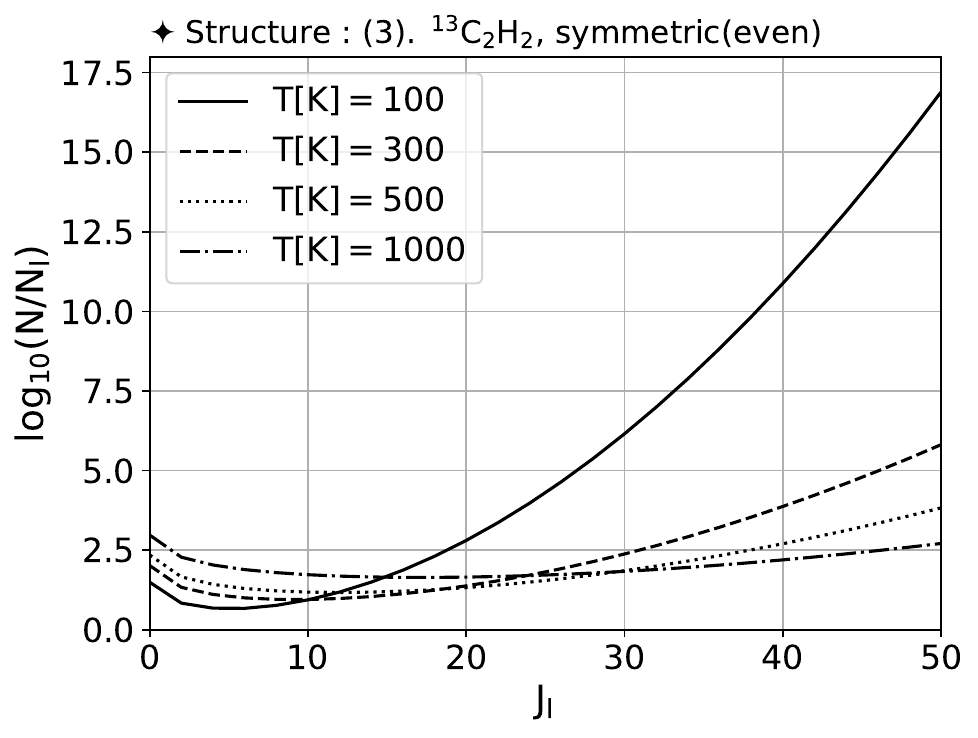}
  \includegraphics[width=0.35\textwidth]{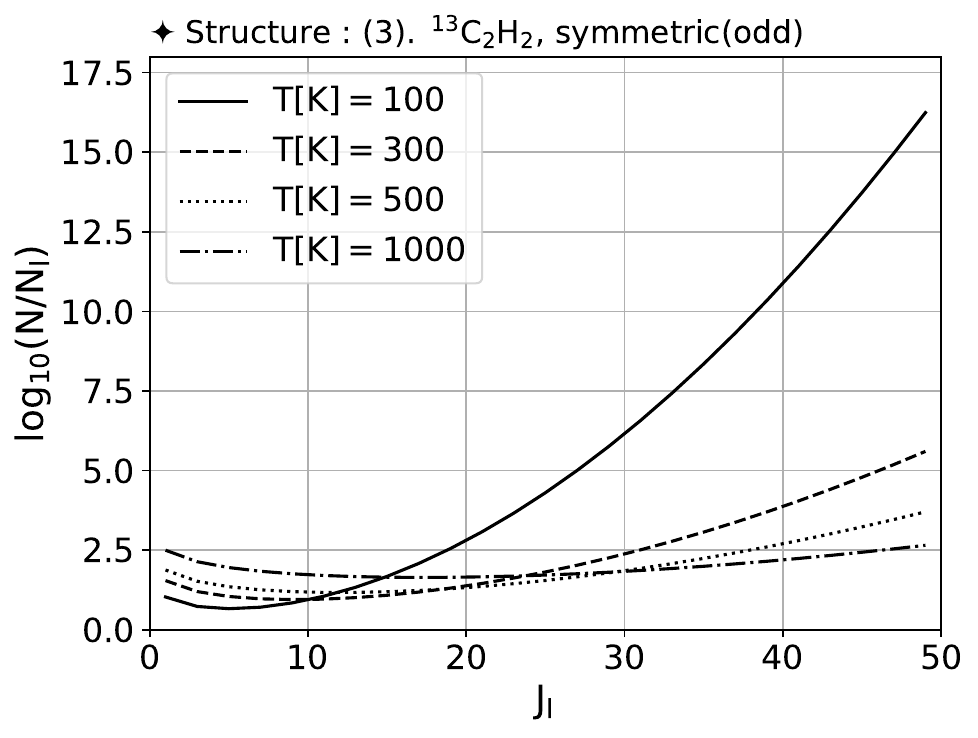}
  \includegraphics[width=0.35\textwidth]{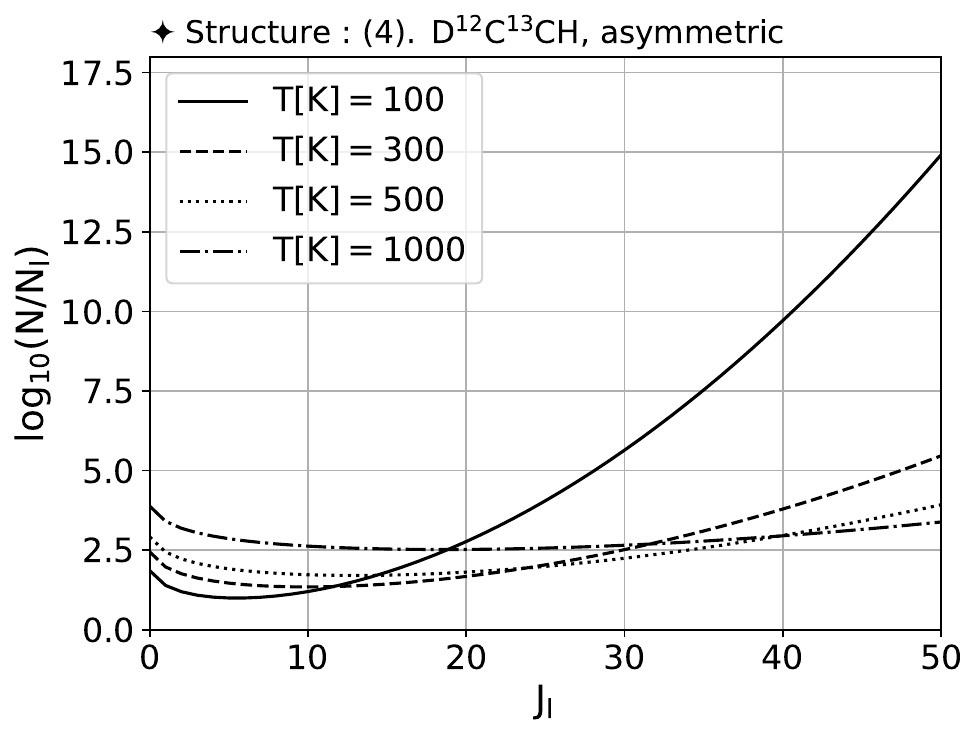}
  \includegraphics[width=0.35\textwidth]{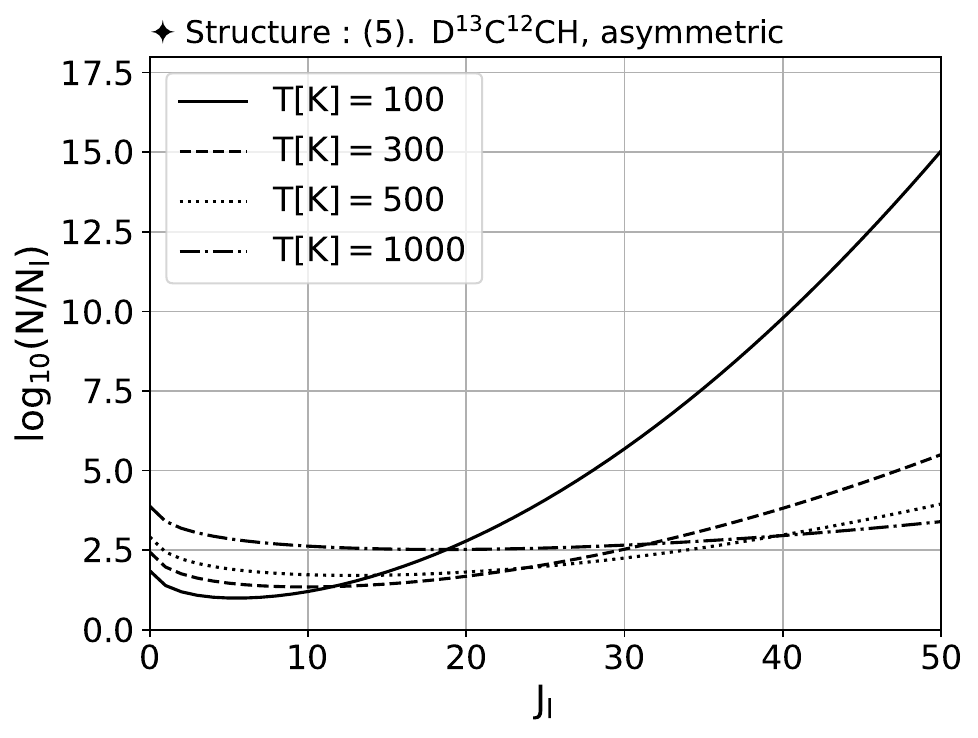}
  \includegraphics[width=0.35\textwidth]{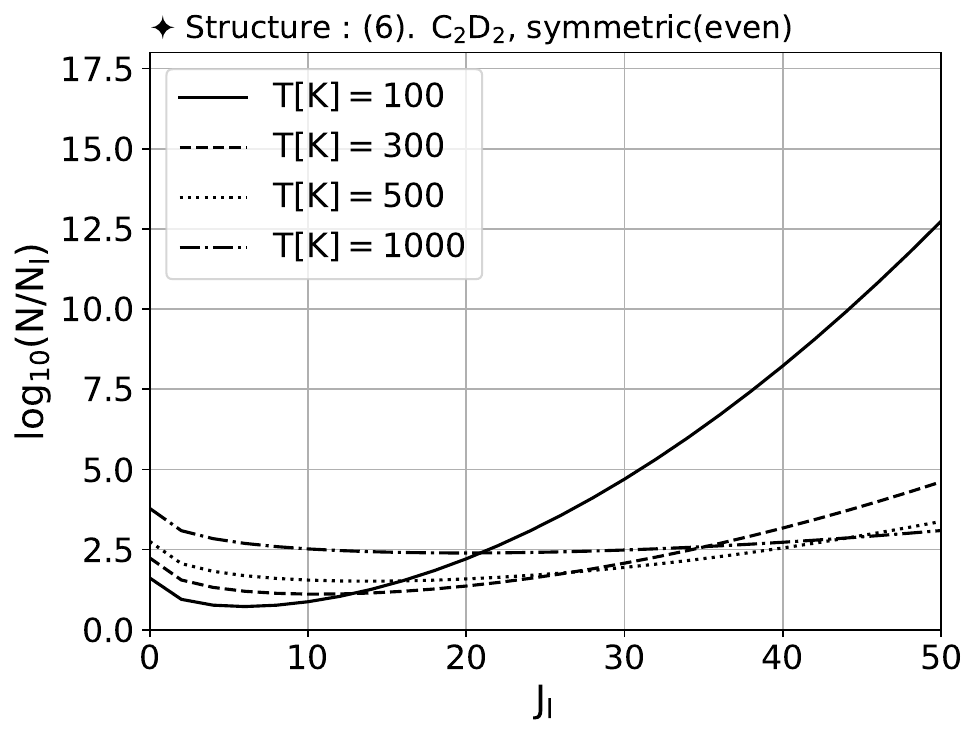}
  \includegraphics[width=0.35\textwidth]{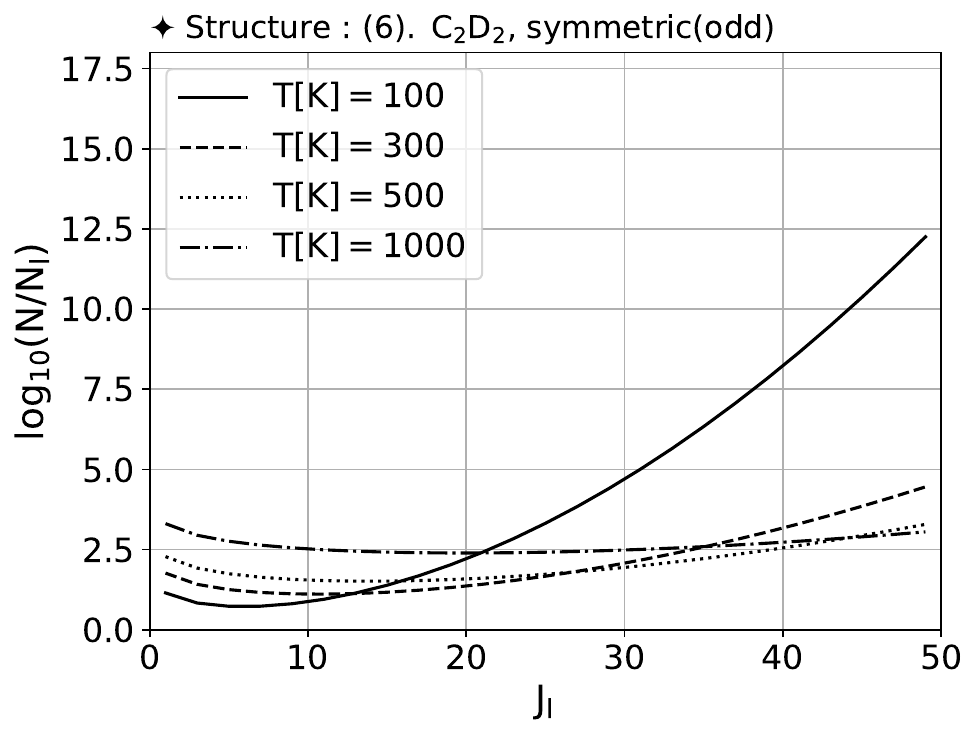}
\caption{
Predicted $\log_{10}(N_l/g_l)$ values for various isotopologues of acetylene, calculated as a function of the lower rotational quantum number $J_l$ at four temperatures $T = 100$(solid), $300$(dashed), $500$(dotted), and $1000$(dot-dashed lines) K. The top row corresponds to $^{13}$C$_2$H$_2$ (molecule ``3''), with separate plots for even (left) and odd (right) rotational states. The middle row shows results for D$^{12}$C$^{13}$CH (molecule ``4'') and D$^{13}$C$^{12}$CH (molecule ``5''), both of which are asymmetric molecules. The bottom row corresponds to $^{12}$C$_2$D$_2$ (molecule ``6''), with separate plots for even (left) and odd (right) rotational states.
The curves are based on partition functions and energy spectra derived from quantum chemical calculations. Even states start at $J_l = 0$, while odd states start at $J_l = 1$, reflecting the symmetry properties of the isotopologues. These plots illustrate the expected population distributions across a range of temperatures, providing insights into the ro-vibrational characteristics of these acetylene isotopologues.
}
\label{fig:others_1}
\end{figure*}

\begin{figure*}[t!]
  \centering
  \includegraphics[width=0.35\textwidth]{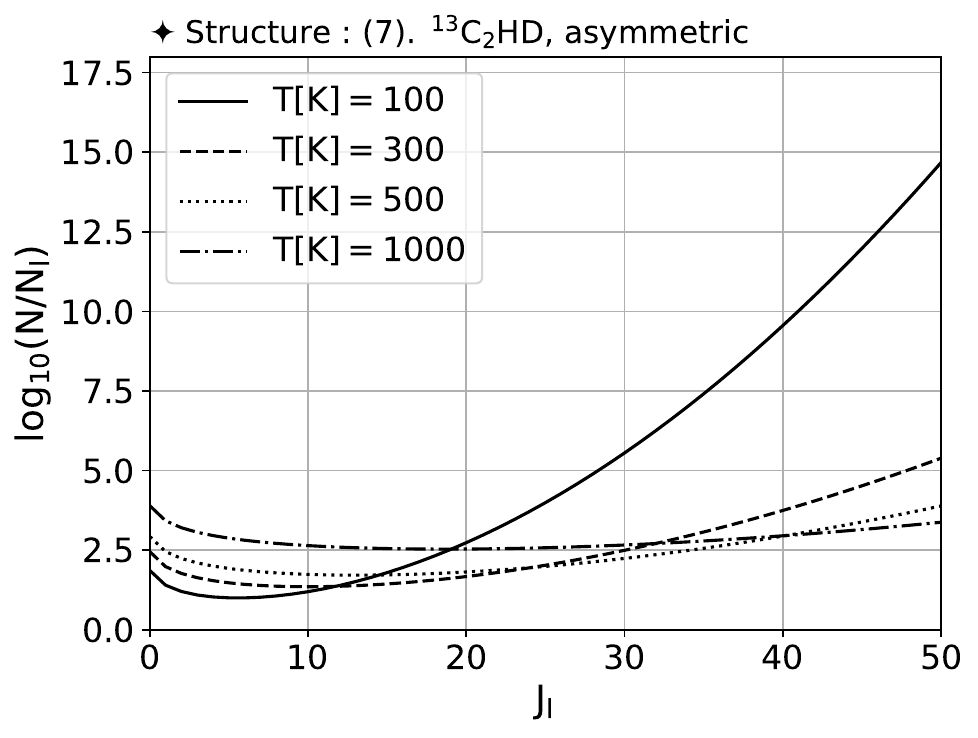}
  \includegraphics[width=0.35\textwidth]{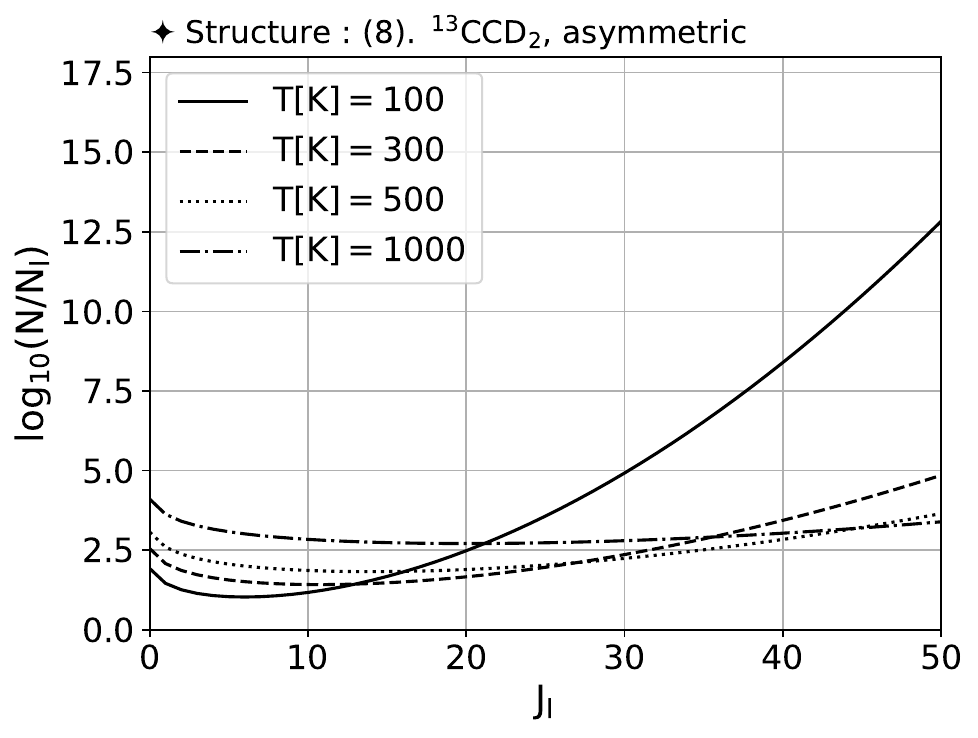}
  \includegraphics[width=0.35\textwidth]{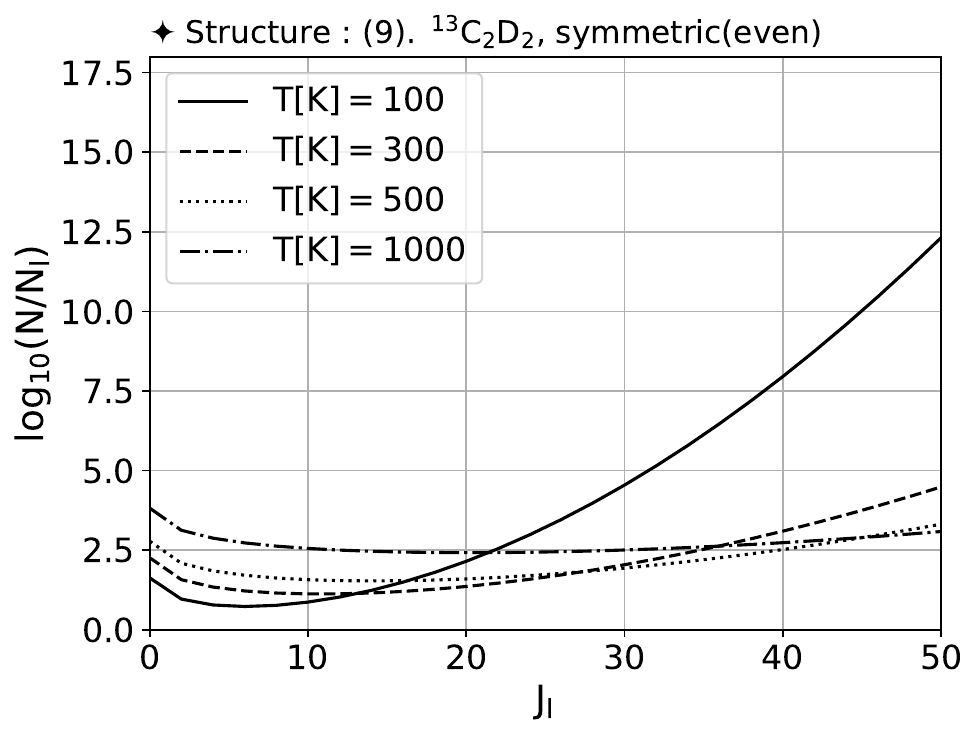}
  \includegraphics[width=0.35\textwidth]{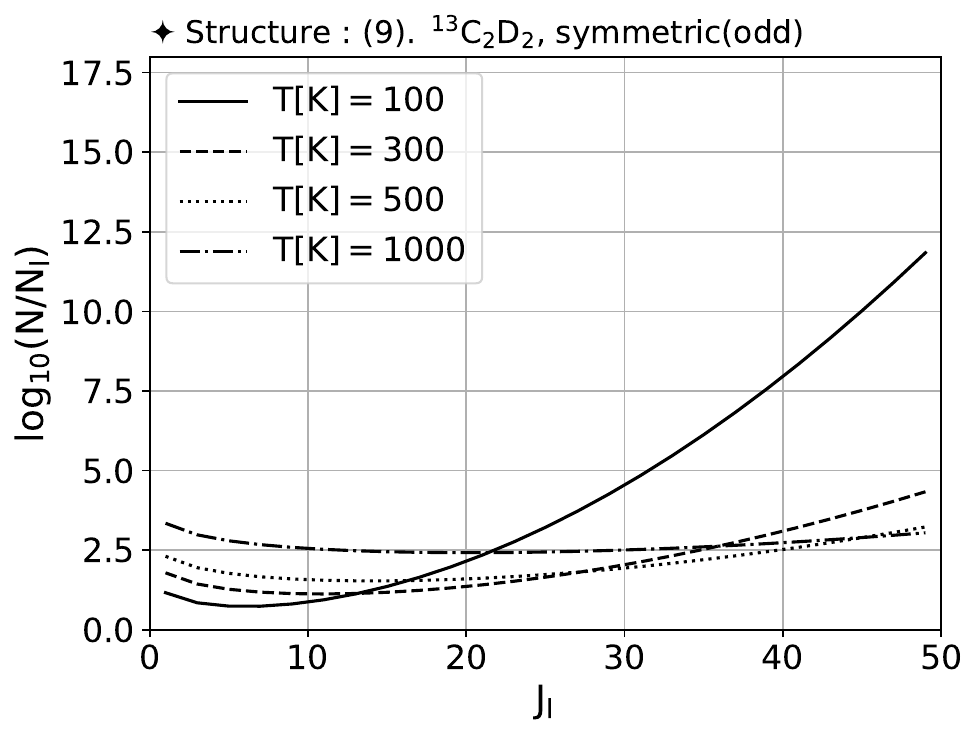}
\caption{
Same as Figure \ref{fig:others_1}, but for the remaining isotopologues of acetylene. The top row shows results for asymmetric molecules: D$^{13}$C$^{13}$CH (molecule ``7'') on the left and D$^{13}$C$^{12}$CH (molecule ``8'') on the right. The bottom row corresponds to symmetric molecules $^{13}$C$^{13}$CD$_2$ (molecule ``9''), with even states on the left and odd states on the right.
} \label{fig:others_2}
\end{figure*}

Figures \ref{fig:others_1} and \ref{fig:others_2} show the predicted $\log_{10}(N_l/g_l)$ values for the remaining isotopologues of acetylene, based on quantum chemical calculations.
Thery includes results for $^{13}$C$_2$H$_2$ (molecule ``3''), D$^{12}$C$^{13}$CH (molecule ``4''), D$^{13}$C$^{12}$CH (molecule ``5''), and $^{12}$C$_2$D$_2$ (molecule ``6''). Figure \ref{fig:others_2} extends the analysis to D$^{13}$C$^{13}$CH (molecule ``7''), D$^{13}$C$^{12}$CH (molecule ``8''), and $^{13}$C$_2$D$_2$ (molecule ``9''). For symmetric molecules ($^{13}$C$_2$H$_2$, $^{12}$C$_2$D$_2$, and $^{13}$C$_2$D$_2$), even $J_l$ states start at $J_l = 0$, while odd $J_l$ states begin at $J_l = 1$, reflecting their symmetry constraints. Asymmetric molecules, such as D$^{12}$C$^{13}$CH and D$^{13}$C$^{12}$CH, exhibit continuous distributions across $J_l$. 
Each panel illustrates how the population distributions vary under different temperatures, providing insights into the statistical and symmetry properties of these isotopologues. These results serve as a theoretical reference for identifying and characterizing acetylene isotopologues in various astrophysical environments.

\subsection{Comparison of results and the $\mathrm{^{12}C / ^{13}C}$ Column Density Ratios}\label{subsec:ratio_comparison}
\subsubsection{Comparison of results}
\begin{table*}[th!]
\caption{
Comparison of estimated temperature (\(T\)) and column density (\(N\)) values for blue and red clump species using two methods: the traditional rotation diagram (RD) method and the \(\chi^2\) fitting method. The \(\chi^2\) method significantly reduces error bars for C\(_2\)H\(_2\) in the blue clump, while revealing larger uncertainties for \(^{13}\)CCH\(_2\) in the red clump due to sparse observational data.
} \label{tab:clump_comparison}
\vspace{0.2cm} 
\centering
\begin{tabular}{llllc}
\hline
\textbf{Species} & \textbf{Source} & \textbf{Estimated Temperature} & \textbf{Estimated Column Density} & \textbf{Method}  \\ 
                 &                 & \textbf{T (K)}                 & \textbf{N (cm$^{-2}$)}            &  \\ 
\hline
\hline
\multicolumn{5}{c}{\textbf{Blue Clump}}                                                         \\ 
\hline
\hline
C$_2$H$_2$ (even, Para) & Nickerson et al. (2023)      & $145 \pm 9$                 & $(1.23 \pm 0.15) \times 10^{16}$             & RD \\ 
                        & This work      & $161^{+3}_{-3}$             & $(0.983^{+0.019}_{-0.018}) \times 10^{16}$   & $\chi^2$ \\ 
\hline
C$_2$H$_2$ (odd, Ortho) & Nickerson et al. (2023)      & $175 \pm 12$                & $(1.50 \pm 0.15) \times 10^{16}$    & RD \\ 
                        & This work      & $192^{+3}_{-3}$             & $(1.31^{+0.015}_{-0.014}) \times 10^{16}$ & $\chi^2$ \\ 
\hline
$^{13}$CCH$_2$          & Nickerson et al. (2023)      & $91 \pm 9$                  & $(2.56 \pm 0.18) \times 10^{15}$   & RD \\ 
                        & This work      & $87.1^{+12.3}_{-10.4}$      & $(2.45^{+0.20}_{-0.19}) \times 10^{15}$ & $\chi^2$ \\ 
\hline
\hline
\multicolumn{5}{c}{\textbf{Red Clump}}                                                         \\ 
\hline
\hline
C$_2$H$_2$ (even, Para) & Nickerson et al. (2023)      & $158 \pm 16$               & $(3.09 \pm 0.57) \times 10^{15}$   & RD \\ 
                        & This work      & $170^{+11}_{-10}$          & $(2.43^{+0.18}_{-0.18}) \times 10^{15}$  & $\chi^2$ \\ 
\hline
C$_2$H$_2$ (odd, Ortho) & Nickerson et al. (2023)      & $229 \pm 27$               & $(3.58 \pm 0.71) \times 10^{15}$   & RD \\ 
                        & This work      & $270^{+21}_{-19}$          & $(2.65^{+0.17}_{-0.16}) \times 10^{15}$  & $\chi^2$ \\ 
\hline
$^{13}$CCH$_2$          & Nickerson et al. (2023)      & $64 \pm 6$                 & $(6.74 \pm 0.64) \times 10^{14}$   & RD \\ 
                        & This work      & $64.4^{+70.7}_{-23.4}$     & $(6.69^{+1.98}_{-1.66}) \times 10^{14}$  & $\chi^2$ \\ 
\hline
\end{tabular}
\begin{flushleft}
\tablerefs{
\citet{nickerson2023mid};
}
\end{flushleft}
\end{table*}

Table \ref{tab:clump_comparison} provides a comparison of the estimated temperature ($T$) and column density ($N$) values for the blue and red clump species, calculated using the traditional rotation diagram (RD) method and the $\chi^2$ fitting method developed in this work. The species include symmetric molecules such as C$_2$H$_2$ (even and odd rotational states) and asymmetric species such as $^{13}$CCH$_2$. The source column distinguishes between the reference values from \citet{nickerson2023mid} and the results obtained in this study, while the method column indicates whether the values were derived using the RD or $\chi^2$ method.

For the blue clump, the $\chi^2$ method provides temperature and column density estimates that differ significantly from the reference results for C$_2$H$_2$. For both the even (Para) and odd (Ortho) states, the center values of T and N derived using the $\chi^2$ method lie outside the 1$\sigma$ uncertainties of the RD method. Additionally, the error bars in our results are substantially smaller, indicating greater precision. This reduction in uncertainties is particularly evident for the odd state, where our error bar for $T$ is reduced by nearly a factor of two. These improvements demonstrate the robustness of the $\chi^2$ method in handling rotational transitions with varying statistical weights.
In the case of $^{13}$CCH$_2$, our results show good agreement with the reference values within 1$\sigma$ uncertainties, suggesting consistency between the two methods. However, the error bars in our results are notably smaller, underscoring the improved reliability of the $\chi^2$ method. This is particularly important for isotopologues with fewer transitions, where precise parameter estimation can be challenging using traditional approaches.

For the red clump, the differences between the two methods are even more pronounced for C$_2$H$_2$. Both even and odd rotational states exhibit center values of $T$ and $N$ that deviate significantly from the reference results, lying well outside the 1$\sigma$ uncertainties of the RD method. For example, the estimated T for the odd state increases from $229 \pm 27$ K (RD) to $270^{+21}_{-19}$ K ($\chi^2$), with a corresponding reduction in uncertainty. Similarly, the column density for the even state decreases significantly, accompanied by a tighter confidence interval. These results highlight the ability of the $\chi^2$ method to account for systematic effects and better constrain physical parameters.
For $^{13}$CCH$_2$, the $\chi^2$-based results for the red clump reveal significantly larger uncertainties compared to the RD method. This discrepancy arises primarily from the limited number of observational data points (four) available for fitting. 
This effect is clearly visible in the left panel of Figure \ref{fig:comp_RD_and_chi2_red_clump}, where the red band representing the uncertainty range from the $\chi^2$ method is significantly wider than the blue band corresponding to the RD method. 
Despite these larger error bars, the center values obtained from both the $\chi^2$ and RD methods are remarkably consistent, falling well within each other's uncertainty ranges. 

\subsubsection{ $\mathrm{^{12}C / ^{13}C}$ Column Density Ratios}
\begin{table*}[th!]
\centering
\caption{Comparison of $^{12}$C/$^{13}$C isotopic ratios derived using the traditional Rotation Diagram (RD) method and the $\chi^2$ fitting method for the blue and red clumps in the Orion IRc2 region. For the red clump, the $\chi^2$ (12C) + RD (13C) approach was used, where the $\chi^2$ fitting method determined the $^{12}$C column density, and the RD method was applied to the $^{13}$C isotopologue due to limited observational data. The differences and percentage differences between the methods are also presented.}
\label{tab:C_ratio_comparison_transposed_with_method}
\begin{tabular}{l|cc|cc}
\textbf{Source $\backslash$ Clump}  & \textbf{Blue Clump}        & \textbf{Method} & \textbf{Red Clump}         & \textbf{Method} \\ \hline
Nickerson et al. (2023) ($^{12}$C/$^{13}$C)       & $21.3 \pm 2.2$             & RD              & $19.8 \pm 3.4$             & RD              \\ 
This Work ($^{12}$C/$^{13}$C)       & $18.72^{+1.54}_{-1.46}$    & $\chi^2$        & $15.07^{+1.61}_{-1.60}$    & $\chi^2$ ($^{12}$C) + RD ($^{13}$C)   \\ \hline
Difference                 & $-2.58^{-0.66}_{-0.74}$    & ---             & $-4.73^{-1.79}_{-1.80}$     & ---             \\
Percent Difference (\%)    & $-12.1^{-30.0}_{-33.6}$    & ---             & $-23.9^{-52.6}_{-52.9}$    & ---             \\ \hline
\end{tabular}
\begin{flushleft}
\tablerefs{
\citet{nickerson2023mid};
}
\end{flushleft}
\end{table*}
This section presents a comparison of $^{12}$C/$^{13}$C ratios derived using the traditional RD method and the $\chi^2$ fitting method, highlighting the advantages of the latter. The $^{12}$C/$^{13}$C ratios were calculated using the formula:

\begin{equation}
\frac{^{12}\mathrm{C}}{^{13}\mathrm{C}} = 
\frac{2\left( 
N(\mathrm{Ortho-C_2H_2})
+ N(\mathrm{Para-C_2H_2})\right)}{N(^{13}\mathrm{CCH_2})},
\end{equation}
where $N(\mathrm{Ortho-C_2H_2})$ and $N(\mathrm{Para-C_2H_2})$ represent the column densities of the ortho and para states of symmetric acetylene, and $N(^{13}\mathrm{CCH_2})$ is the column density of the $^{13}$C isotopologue. This formula accounts for molecular symmetry and statistical weights, ensuring consistency and accuracy in the isotopic ratio determination.

Table~\ref{tab:C_ratio_comparison_transposed_with_method} presents a detailed comparison of the $^{12}$C/$^{13}$C isotopic ratios derived for the blue and red clumps in the Orion IRc2 region using two different methods: the traditional Rotation Diagram (RD) method and the $\chi^2$ fitting method introduced in this study. The table also includes the absolute differences and percentage differences between the results of the two methods, highlighting the systematic biases inherent in the RD approach and the improvements achieved with the $\chi^2$ method.

For the blue clump, the $\chi^2$ method yields a $^{12}$C/$^{13}$C ratio of $18.72^{+1.54}_{-1.46}$, which is approximately 12.1\% lower than the RD-derived value of $21.3 \pm 2.2$. This difference reflects the systematic overestimation of the RD method due to its reliance on linear extrapolation and the separate fitting of temperature ($T$) and column density ($N$). Additionally, the $\chi^2$ method reduces the uncertainty of the isotopic ratio by nearly 30\%, demonstrating its superior precision and reliability.

For the red clump, a hybrid approach, $\chi^2$ ($^{12}$C) + RD ($^{13}$C), was employed. In this approach, the $\chi^2$ method was used to determine the $^{12}$C column density, while the RD method was applied to the $^{13}$C isotopologue due to the limited number of observational data points. This combination yields a $^{12}$C/$^{13}$C ratio of $15.07^{+1.61}_{-1.60}$, which is 23.9\% lower than the RD-derived value of $19.8 \pm 3.4$. While the $\chi^2$ method improves the overall accuracy for $^{12}$C, the use of the RD method for $^{13}$C contributes to slightly larger uncertainties compared to the blue clump.

The differences in the derived ratios are significant and underscore the importance of adopting the $\chi^2$ fitting method, which minimizes biases by simultaneously fitting $T$ and $N$, incorporating statistical weights, and accounting for degeneracy factors. This method ensures more robust and accurate determination of isotopic ratios, particularly for datasets with adequate observational coverage.

Despite the limitations for the red clump, the results for both clumps confirm the systematic overestimation of $^{12}$C/$^{13}$C ratios by the RD method, which could impact astrophysical interpretations of isotopic fractionation processes. The hybrid $\chi^2$ ($^{12}$C) + RD ($^{13}$C) approach provides a practical solution when data for $^{13}$C is sparse, but its limitations highlight the need for more extensive observational data to fully leverage the advantages of the $\chi^2$ method.

\section{Conclusions}\label{sec:conclusions}
In this study, we conducted a comprehensive theoretical investigation of the infrared spectra of acetylene and its isotopologues, focusing on the effects of isotopic substitution involving $^{13}$C and deuterium (D). Through extensive quantum chemical calculations using methods such as Hartree-Fock (HF), Density Functional Theory (DFT), and Møller-Plesset perturbation theory (MP2), we determined rotational constants and centrifugal distortion constants ($\tilde{D}_J$) for molecules lacking observational data. This rigorous approach allowed us to systematically account for all possible isotopologues of acetylene, including combinations of $^{13}$C and D substitution. While $^{13}$C substitution induces subtle vibrational shifts on the order of a few cm$^{-1}$, detectable only with high-precision instruments like SOFIA, D substitution causes significant vibrational shifts of tens of cm$^{-1}$ due to its larger impact on the reduced mass. These differences in sensitivity and detectability underscore the importance of accurate rotational and vibrational constants in studying isotopic effects across a range of astrophysical conditions (for more details on quantum chemical computations, see Appendices \ref{apd:QCMs} and \ref{apd:IR_spectra}).

Comparisons with observational data from instruments such as SOFIA validated our computational approaches. The integration of the HITRAN database and the application of Doppler and Local Standard of Rest (LSR) adjustments ensured the accuracy of our theoretical models. These comparisons highlighted the value of isotopologue-specific spectra in deciphering the molecular composition of astrophysical environments like the Orion Nebula, as well as their critical role in tracing stellar nucleosynthesis and the chemical evolution of galaxies.

A significant contribution of this study lies in the application of the $\chi^2$ fitting method, which effectively addresses the limitations of the traditional RD approach. By simultaneously fitting temperature ($T$) and column density ($N$) while incorporating statistical weights, transition strengths, and degeneracy factors, the $\chi^2$ method minimizes systematic biases and improves the reliability of derived parameters. 
For the blue clump, the $\chi^2$ method achieved higher precision, reducing uncertainties in $T$ and $N$ for both even and odd rotational states of C$_2$H$_2$, while providing results consistent with or slightly adjusted from the RD method. Similarly, for $^{13}$CCH$_2$, the $\chi^2$ method yielded column densities that aligned closely with the RD results, while significantly reducing the associated uncertainties.
In the red clump, the $\chi^2$ method improved precision for C$_2$H$_2$, showing higher temperatures and slightly lower column densities compared to the RD method, accompanied by smaller error margins. However, for $^{13}$CCH$_2$, the limited number of observational data points led to increased uncertainties in the $\chi^2$ method's results, despite the central values being consistent with the RD method. This highlights a key limitation of the $\chi^2$ approach when applied to datasets with very few observations, as the increased number of fitting parameters exacerbates the impact of limited data on the robustness of the fit.
Overall, these findings demonstrate the robustness and precision of the $\chi^2$ method for deriving physical parameters from molecular spectra, particularly for well-sampled datasets, while emphasizing the need for sufficient observational data to fully exploit its advantages in future studies.

Through the $\chi^2$ method, we identified systematic overestimations of $^{12}$C/$^{13}$C ratios by the RD method: approximately 12.1\% for the blue clump and 23.9\% for the red clump. For the blue clump, the $\chi^2$ method yielded a ratio of $18.72^{+1.54}_{-1.46}$ compared to $21.3 \pm 2.2$ from the RD method, with uncertainties reduced by nearly 30\%. This improvement demonstrates the robustness of the $\chi^2$ method in mitigating biases and enhancing precision. 
For the red clump, the $\chi^2$ method estimated a ratio of $15.07^{+1.61}_{-1.60}$, significantly lower than the RD-derived value of $19.8 \pm 3.4$. However, the uncertainty for $^{13}$CCH$_2$ was notably larger due to the limited observational data, as explained earlier. To address this, a hybrid approach combining $\chi^2$ fitting for C$_2$H$_2$ and the RD method for $^{13}$CCH$_2$ was employed, ensuring reliable results while acknowledging the limitations of the $\chi^2$ method in sparse datasets. 

In addition to demonstrating the accuracy of the $\chi^2$ fitting method, this study introduces a Python package designed to automate the spectral analysis process. The package integrates quantum chemical calculations and robust fitting techniques, enabling precise analyses of acetylene isotopologues. It supports both RD and $\chi^2$ fitting methods, providing direct comparisons and facilitating validation of results.
Freely available on GitHub, the package includes precomputed rotational constants, partition functions (TIPs), and vibrational-rotational spectra for a wide range of isotopologues, addressing gaps in existing databases like HITRAN. For details on the package's implementation and usage, including applications to new observational data, refer to Appendix \ref{apd:TOPSEGI}.

In summary, this study highlights the significant advantages of the $\chi^2$ fitting method for analyzing complex molecular spectra. By reducing uncertainties and mitigating systematic biases, the method provides a robust and reliable framework for deriving isotopic ratios and molecular abundances, offering critical insights into the physical conditions of astrophysical environments. The observed systematic overestimations in isotopic ratios by the RD method underscore the need to revisit earlier studies to ensure accurate interpretations of stellar nucleosynthesis and galactic chemical evolution. 
Looking forward, extending the $\chi^2$ methodology to other isotopologues and molecular systems will enable more comprehensive explorations of interstellar chemistry and the mechanisms driving galactic evolution. This study not only advances our understanding of isotopic fractionation but also establishes a foundation for future high-precision spectroscopic analyses, reinforcing the importance of adopting modern, data-driven approaches in astrophysical research.

\section*{Acknowledgments}
This work was supported by the National Research Foundation of Korea (NRF) grant funded by the Korea government (RS-2021-NR060129), and we would like to express our sincere gratitude to Dr. Naseem Rangwala and Dr. Sarah Nickerson for their invaluable advice on the analysis methods for acetylene observation data. Their insights and suggestions have greatly contributed to the development and completion of this work.
\newpage
\appendix
\twocolumngrid
\section{Corrections for Doppler shifts and Local Standard of Rest (LSR) velocities}

To accurately interpret high-resolution infrared data, such as those obtained from SOFIA \citep{rangwala2018high}, corrections considering both the Doppler effect and Local Standard of Rest (LSR) velocities are essential. The LSR is a coordinate system defined based on the average motion of stars near the Sun. The International Astronomical Union (IAU) defines the LSR, based on the B1900 equinox, at right ascension ($\alpha \mathrm{= 18h \; 00m}$),  declination ($\delta \mathrm{= +30^{\circ}}$) with a velocity of 20km/s. However, modern research uses the J2000 standard, requiring a transformation from B1900 coordinates to J2000 for analysis.\\
To perform LSR corrections, the Julian Date (JD) at the time of observation and the target's RA and DEC are required to calculate the line-of-sight velocity difference between the geocentric frame and the LSR frame. This involves calculating the velocity difference between the LSR and the heliocentric (Sun-centered) frame based on the defined LSR standard, adding the velocity difference between the geocentric frame and the heliocentric frame at the observation time, and combining these two differences to determine the final line-of-sight velocity between the geocentric and LSR frames.\\
For the observational data of the Orion IRc2 region used in Figure \ref{fig:reproduce}, the observation duration was relatively short at approximately 28 minutes, allowing the velocity variations over the observation period to be neglected. The Doppler effect due to the LSR was calculated to be approximately 7.29 km/s for the observational data, and Figure \ref{fig:reproduce} presents the results calculated based on the start time of the observation.\\
Such LSR corrections accurately account for the Doppler effects present in the observational data, providing the essential precision required for molecular spectrum analysis. This enables the precise determination of the observed molecular spectral line positions and facilitates the quantitative derivation of physical parameters.

\section{Quantum Chemical Calculation Methods}\label{apd:QCMs}
In this study, quantum chemical calculations were performed using \texttt{Gaussian16} \citep{foresman1996exploring} with various computational methods to ensure the accuracy and reliability of the results. Specifically, the \textit{Hartree-Fock (HF)} \citep{helgaker2013molecular} method was employed for its simplicity and computational speed, providing a foundational understanding of the electronic structure. \textit{Density Functional Theory (DFT)} with the \textit{B3LYP} \citep{PhysRevA.38.3098, PhysRevB.37.785} and \textit{PBE0} \citep{PhysRevLett.77.3865} functionals was chosen for its balance between accuracy and computational efficiency, offering improved treatment of electron correlation effects compared to HF. Additionally, the \textit{second-order M{\o}ller-Plesset perturbation theory (MP2) \citep{PhysRev.46.618}} method was utilized to achieve higher precision in accounting for electron correlation.

Geometric optimizations were carried out using the \texttt{6-31+G*} basis set, which includes polarization and diffuse functions to accurately describe molecular geometries and electronic distributions. This basis set was selected for its ability to capture the subtle effects of isotopic substitution on molecular properties.

Vibrational frequency calculations were performed for acetylene and its isotopologues, including isotopically substituted species. To evaluate the impact of isotopic substitutions, specific atoms were systematically replaced (e.g., hydrogen with deuterium or carbon-12 with carbon-13), followed by reoptimization of molecular geometries and recalculation of vibrational spectra. The changes in reduced mass due to isotopic substitution were explicitly considered to analyze their influence on vibrational frequencies.

Among the tested methods, \textit{B3LYP} showed the best agreement with experimental data for most vibrational and rotational parameters, making it the primary method employed in this study. Other methods, such as HF, PBE0, and MP2, exhibited lower consistency with experimental results and were therefore not selected for further analysis.

\section{IR Spectra and Rotational Constants of Acetylene Isotopes}\label{apd:IR_spectra}
We present the vibrational frequencies (\(\nu_1 \sim \nu_5\)) and rotational constants (\(\tilde{B}\)) for acetylene (\(^{12}\)C\(_2\)H\(_2\)) and its isotopologues (\(^{13}\)C- and deuterium-substituted variants), calculated using \textit{HF}, \textit{B3LYP}, \textit{PBE0}, and \textit{MP2} methods. Table \ref{tab:quantum_chemical_models_Btilde} summarizes these results, with infrared-active modes in black and inactive modes in gray. Substituted \(^{13}\)C or D atoms are marked with red circles to facilitate isotopologue identification. This dataset is essential for modeling isotopologues, including those not yet observed, and understanding their roles in astrophysical environments.

A frequency scaling factor (FSF) of 1.0000 was used initially for uniform comparison. For isotopologues with experimental data, the FSF was iteratively adjusted to improve agreement with observations, as detailed in Section \ref{subsubsec:TIPS}.

These parameters support the calculation of total internal partition sums (TIPS), which are critical for determining molecular column densities and isotopic ratios. Rotational constants (\(\tilde{B}\)) were validated against experimental data, demonstrating high accuracy for symmetric and asymmetric isotopologues. The results have been incorporated into the Python package developed for this study, enabling automated spectral modeling and high-resolution analyses.

\twocolumngrid
\begin{deluxetable*}{ccrrrrrrr}
\tablecaption{
Infrared (IR) vibrational frequencies (\(\nu_1\), \(\nu_2\), \(\nu_3\), \(\nu_4\), \(\nu_5\)) and rotational constants (\(\tilde{B}\)) for acetylene isotopologues (\(^{12}\)C\(_2\)H\(_2\), \(^{13}\)C-substituted, and deuterium-substituted variants) calculated using various quantum chemical methods: Hartree-Fock (HF), B3LYP, PBE0, and MP2. The ``Structure'' column visually distinguishes the isotopologues, with substituted \(^{13}\)C and deuterium (D) atoms marked by red circles. Vibrational frequencies shown in black indicate infrared-active modes, while gray denotes infrared-inactive modes. A frequency scaling factor (FSF) of 1.0000 was used for all calculations; however, users can adjust the FSF as needed to align the theoretically computed vibrational frequencies more closely with experimental values. These results, detailed further in Section \ref{apd:IR_spectra}, are essential for accurate spectral modeling and provide key data for isotopologues not covered in standard databases such as HITRAN.
}
\label{tab:quantum_chemical_models_Btilde}
\tablehead{
    \colhead{Molecule} & \colhead{Structure} & \colhead{Method} & \colhead{$\nu_{1}$} & \colhead{$\nu_{2}$} & \colhead{$\nu_{3}$} & \colhead{$\nu_{4}$} & \colhead{$\nu_{5}$} & \colhead{$\mathrm{\tilde{B}}$} \\
    \colhead{ID} & \colhead{} & \colhead{FSF = 1.0000} & \colhead{$\mathrm{[cm^{-1}]}$} & \colhead{$\mathrm{[cm^{-1}]}$} & \colhead{$\mathrm{[cm^{-1}]}$} & \colhead{$\mathrm{[cm^{-1}]}$} & \colhead{$\mathrm{[cm^{-1}]}$} & \colhead{$\mathrm{[cm^{-1}]}$} 
}
\startdata
\multirow{4}{*}{0} & 
\multirow{4}{*}{
\begin{minipage}{.3\textwidth}
\centering
\includegraphics[width=3.5cm, keepaspectratio]{./img/Acetylene_vol0.pdf}
\end{minipage}
}
                                        & HF     & \color{gray}{3719.692} & \color{gray}{2246.893} & 3607.484 & \color{gray}{793.610} & 882.237 & 1.2113 \\
                   &                    & B3LYP  & \color{gray}{3542.390} & \color{gray}{2087.886} & 3442.297 & \color{gray}{534.175} & 774.627 & 1.1766 \\
                   &                    & PBE0   & \color{gray}{3563.447} & \color{gray}{2102.532} & 3460.432 & \color{gray}{575.732} & 780.424 & 1.1770 \\
                   &                    & MP2    & \color{gray}{3568.687} & \color{gray}{2001.497} & 3480.443 & \color{gray}{374.534} & 747.863 & 1.1585 \\ \hline
\multirow{4}{*}{1} & 
\multirow{4}{*}{
\begin{minipage}{.3\textwidth}
\centering
\includegraphics[width=3.5cm, keepaspectratio]{./img/Acetylene_vol1.pdf}
\end{minipage}
}
                                        & HF     & 3705.880 & 2211.601 & 3601.439 & 787.455 & 880.998 & 1.1824 \\
                   &                    & B3LYP  & 3529.920 & 2054.704 & 3436.511 & 530.081 & 773.484 & 1.1485 \\
                   &                    & PBE0   & 3550.771 & 2069.187 & 3454.626 & 571.308 & 779.278 & 1.1489 \\
                   &                    & MP2    & 3557.340 & 1969.037 & 3474.553 & 371.702 & 746.750 & 1.1307 \\ \hline                  
\multirow{4}{*}{2} & 
\multirow{4}{*}{
\begin{minipage}{.3\textwidth}
\centering
\includegraphics[width=3.5cm, keepaspectratio]{./img/Acetylene_vol2.pdf}
\end{minipage}
}
                                        & HF     & 3670.580 & 2082.710 & 2843.301 & 657.380 & 837.488 & 1.0184 \\
                   &                    & B3LYP  & 3498.078 & 1946.382 & 2695.795 & 466.499 & 697.315 & 0.9901 \\
                   &                    & PBE0   & 3517.957 & 1958.769 & 2712.453 & 499.230 & 707.655 & 0.9903 \\
                   &                    & MP2    & 3528.849 & 1883.762 & 2696.072 & 332.028 & 662.589 & 0.9762 \\ \hline
\multirow{4}{*}{3} & 
\multirow{4}{*}{
\begin{minipage}{.3\textwidth}
\centering
\includegraphics[width=3.5cm, keepaspectratio]{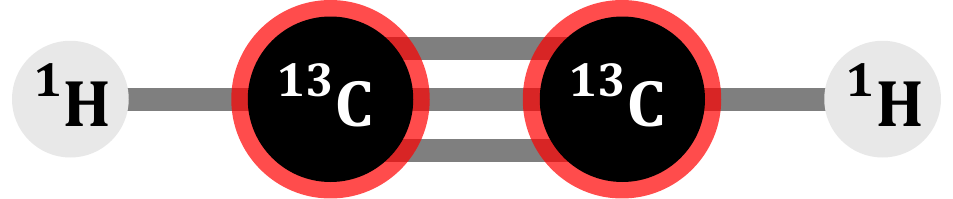}
\end{minipage}
}
                                        & HF     & \color{gray}{3691.056} & \color{gray}{2175.212} & 3596.685 & \color{gray}{781.435} & 879.596 & 1.1529 \\
                   &                    & B3LYP  & \color{gray}{3516.414} & \color{gray}{2020.534} & 3431.992 & \color{gray}{526.002} & 772.308 & 1.1197 \\
                   &                    & PBE0   & \color{gray}{3537.079} & \color{gray}{2034.844} & 3450.073 & \color{gray}{566.906} & 778.088 & 1.1201 \\
                   &                    & MP2    & \color{gray}{3544.806} & \color{gray}{1935.681} & 3470.024 & \color{gray}{368.870} & 745.624 & 1.1022 \\ \hline
\multirow{4}{*}{4} & 
\multirow{4}{*}{
\begin{minipage}{.3\textwidth}
\centering
\includegraphics[width=3.5cm, keepaspectratio]{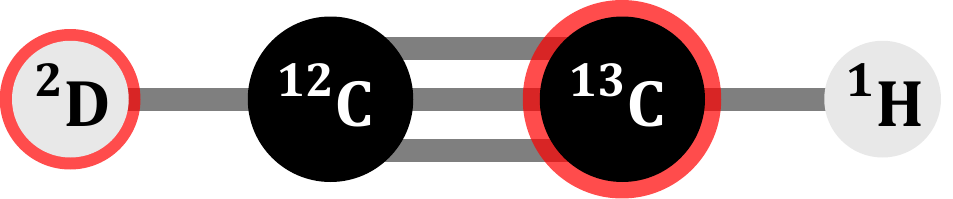}
\end{minipage}
}
                                        & HF     & 3650.710 & 2053.231 & 2836.946 & 655.611 & 831.742 & 0.9935 \\
                   &                    & B3LYP  & 3479.841 & 1917.850 & 2690.606 & 463.780 & 694.732 & 0.9659 \\
                   &                    & PBE0   & 3499.486 & 1930.164 & 2707.181 & 496.565 & 704.675 & 0.9661 \\
                   &                    & MP2    & 3511.667 & 1854.562 & 2692.251 & 329.755 & 660.872 & 0.9522 \\ \hline
\multirow{4}{*}{5} & 
\multirow{4}{*}{
\begin{minipage}{.3\textwidth}
\centering
\includegraphics[width=3.5cm, keepaspectratio]{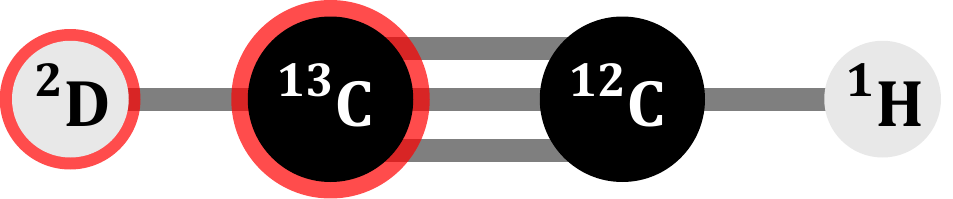}
\end{minipage}
}
                                        & HF     & 3668.247 & 2067.710 & 2803.612 & 648.627 & 837.200 & 1.0018 \\
                   &                    & B3LYP  & 3496.169 & 1930.915 & 2659.920 & 460.335 & 697.020 & 0.9740 \\
                   &                    & PBE0   & 3515.974 & 1943.344 & 2676.210 & 492.517 & 707.511 & 0.9742 \\
                   &                    & MP2    & 3527.424 & 1866.274 & 2663.404 & 327.934 & 661.778 & 0.9601 \\ \hline
\multirow{4}{*}{6} & 
\multirow{4}{*}{
\begin{minipage}{.3\textwidth}
\centering
\includegraphics[width=3.5cm, keepaspectratio]{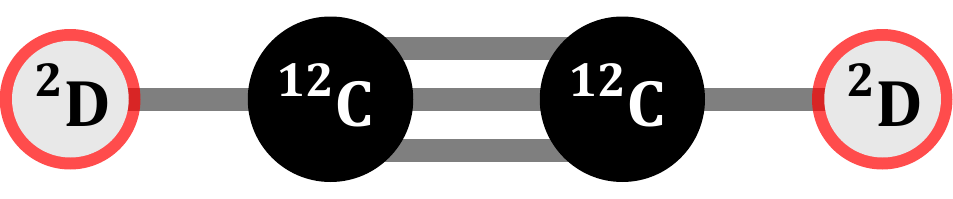}
\end{minipage}
}
                                        & HF     & \color{gray}{2997.782} & \color{gray}{1972.156} & 2648.723 & \color{gray}{647.765} & 662.808 & 0.8692 \\
                   &                    & B3LYP  & \color{gray}{2833.916} & \color{gray}{1846.155} & 2527.438 & \color{gray}{445.952} & 568.754 & 0.8457 \\
                   &                    & PBE0   & \color{gray}{2853.342} & \color{gray}{1857.424} & 2540.753 & \color{gray}{480.778} & 573.011 & 0.8458 \\
                   &                    & MP2    & \color{gray}{2817.727} & \color{gray}{1793.148} & 2555.446 & \color{gray}{312.166} & 549.103 & 0.8346 \\ \hline
\multirow{4}{*}{7} & 
\multirow{4}{*}{
\begin{minipage}{.3\textwidth}
\centering
\includegraphics[width=3.5cm, keepaspectratio]{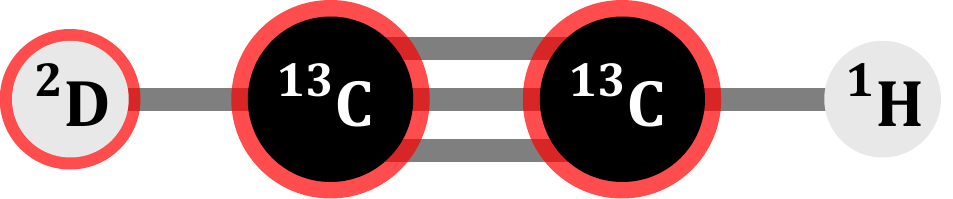}
\end{minipage}
}
                                        & HF     & 3648.706 & 2036.737 & 2797.652 & 646.773 & 831.499 & 0.9764 \\
                   &                    & B3LYP  & 3478.197 & 1901.036 & 2655.100 & 457.617 & 694.411 & 0.9491 \\
                   &                    & PBE0   & 3497.776 & 1913.382 & 2671.303 & 489.848 & 704.507 & 0.9493 \\
                   &                    & MP2    & 3510.435 & 1835.879 & 2659.907 & 325.662 & 660.044 & 0.9355 \\ \hline
\multirow{4}{*}{8} & 
\multirow{4}{*}{
\begin{minipage}{.3\textwidth}
\centering
\includegraphics[width=3.5cm, keepaspectratio]{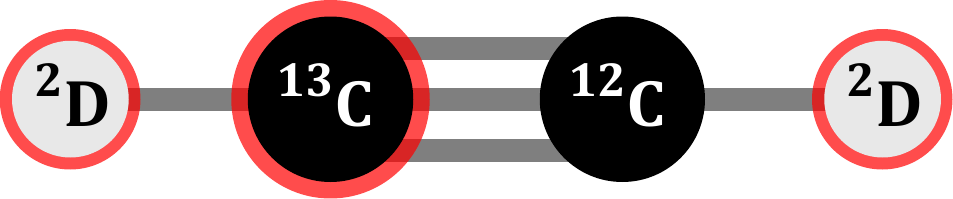}
\end{minipage}
}
                                        & HF     & 2965.050 & 1955.340 & 2640.730 & 644.736 & 656.746 & 0.8542 \\
                   &                    & B3LYP  & 2804.642 & 1829.336 & 2519.795 & 441.000 & 567.229 & 0.8310 \\
                   &                    & PBE0   & 2823.697 & 1840.609 & 2533.075 & 475.411 & 571.500 & 0.8311 \\
                   &                    & MP2    & 2791.728 & 1774.858 & 2547.684 & 308.749 & 547.595 & 0.8201 \\ \hline
\multirow{4}{*}{9} & 
\multirow{4}{*}{
\begin{minipage}{.3\textwidth}
\centering
\includegraphics[width=3.5cm, keepaspectratio]{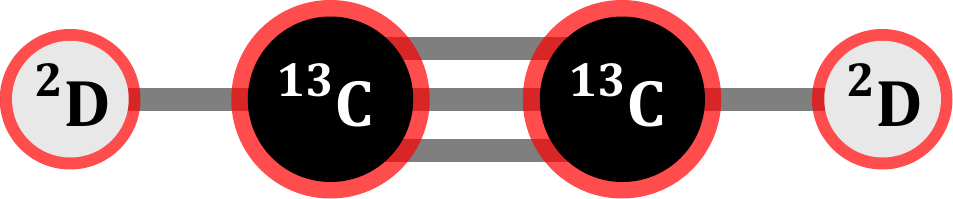}
\end{minipage}
}
                                        & HF     & \color{gray}{2931.432} & \color{gray}{1937.423} & 2633.996 & \color{gray}{644.163} & 648.181 & 0.8387 \\
                   &                    & B3LYP  & \color{gray}{2774.495} & \color{gray}{1811.482} & 2513.385 & \color{gray}{436.132} & 565.592 & 0.8160 \\
                   &                    & PBE0   & \color{gray}{2793.182} & \color{gray}{1822.756} & 2526.626 & \color{gray}{470.176} & 569.825 & 0.8160 \\
                   &                    & MP2    & \color{gray}{2764.762} & \color{gray}{1755.579} & 2541.237 & \color{gray}{305.350} & 546.050 & 0.8050 \\
\enddata
\end{deluxetable*}

\section{TIPS for acetylene isotopologues}\label{apd:TIP}
\begin{figure*}[t!]
  \centering
    \caption{
   Temperature dependence of total internal partition sums (\(q_{\text{tot}}\)) for various isotopologues of acetylene, displayed over a range from 0 to 1000 K. The isotopologues include \(^{13}\text{C}_2\text{H}_2\), \(\text{D}^{12}\text{C}^{13}\text{CH}\), \(\text{D}^{13}\text{C}^{12}\text{CH}\), \(\text{D}^{12}\text{C}^{12}\text{CD}\), \(\text{D}^{13}\text{C}^{13}\text{CH}\), \(\text{D}^{13}\text{C}^{12}\text{CD}\), and \(\text{D}^{13}\text{C}^{13}\text{CD}\). Each curve represents the calculated partition sum for one isotopologue, highlighting the effect of isotopic substitution on the partition function as temperature increases. The plot emphasizes how isotopic differences in hydrogen and carbon atoms impact the thermodynamic properties of acetylene across the specified temperature range.
  } \label{fig:tips_others}
  \includegraphics[width=1.00\textwidth]{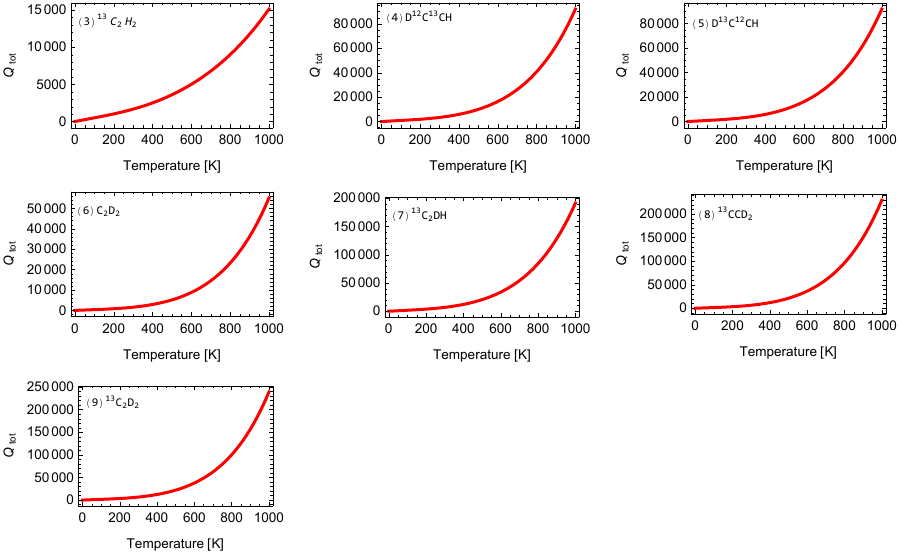}
\end{figure*}

\twocolumngrid
\begin{deluxetable*}{rr|rrrr|rrrr}
\tablecaption{Input data format required for the Python package. The table includes precomputed partition functions (TIPS) with temperature ($T$) and $Q(T)$, theoretical values such as transition branches, rotational quantum numbers ($J_l$), degeneracy factors ($g_l$), and energy levels ($E_l$), as well as observed values including column densities ($N_l$) and their uncertainties ($\sigma_{N_l}$). These inputs allow the package to perform spectral analysis using both the RD and $\chi^2$ fitting methods.}
\label{tab:input_data}
\tablehead{
\multicolumn{2}{c|}{\underline{TIPS}} & \multicolumn{4}{c|}{\underline{Theoretical values}} & \multicolumn{4}{c}{\underline{Observed values}} \\ 
    \colhead{T[K]} & \multicolumn{1}{c|}{Q(T)} & \colhead{Branch} & \colhead{$J_{l}$} & \colhead{$g_{l}$} & \multicolumn{1}{c|}{$\mathrm{E}_{l}\mathrm{[cm^{-1}]}$} & \colhead{Branch}	& \colhead{$J_{l}$}	& \colhead{$\mathrm{N}_{l} \mathrm{[\times 10^{10} cm^{-2}]}$}	& \colhead{$\sigma_{\mathrm{N}_{l}} \mathrm{[\times 10^{10} cm^{-2}]}$}
}
\startdata
1        &    21.0856 & P        & 0        & 15       & 0.0000   & P &  2 &	 2.3100 & 0.2560 \\
2        &    36.8198 & P        & 1        & 63       & 1.6320   & P &  8 &	12.5000 &0.9710  \\
3        &    52.3683 & P        & 2        & 75       & 4.8957   & Q &  6 &	 5.1000 &0.3700  \\
4        &    67.6788 & \scalebox{0.5}{$\vdots$} & \scalebox{0.5}{$\vdots$} & \scalebox{0.5}{$\vdots$} & \scalebox{0.5}{$\vdots$} & Q & 10 &	 6.9500 &0.8500  \\
5        &    82.9553 & Q        & 0        & 15       & 0.0000   & R &  0 &	 0.4560 &0.0455  \\
6        &    98.2421 & Q        & 1        & 63       & 1.6320   & R &  2 &	 4.0000 &0.2720  \\  
7        &   113.5415 & Q        & 2        & 75       & 4.8957   & R &  4 &	 8.3800 &0.8540  \\
8        &   128.8501 & \scalebox{0.5}{$\vdots$} & \scalebox{0.5}{$\vdots$} & \scalebox{0.5}{$\vdots$} & \scalebox{0.5}{$\vdots$} & R &  6 &	 9.7800 &0.8940  \\
9        &   144.1653 & R        & 0        & 15       & 0.0000   & R &  8 &	 9.7800 &0.7870  \\
10       &   159.4852 & R        & 1        & 63       & 1.6320   & R & 10 &	10.3000 &1.0200  \\
11       &   174.8086 & R        & 2        & 75       & 4.8957   & R & 16 &	 5.4800 &0.5930  \\
\scalebox{0.5}{$\vdots$} &  \scalebox{0.5}{$\vdots$}  & \scalebox{0.5}{$\vdots$} & \scalebox{0.5}{$\vdots$} & \scalebox{0.5}{$\vdots$} & \scalebox{0.5}{$\vdots$} & R & 28 &	 1.8600 &0.1570
\enddata
\tablecomments{The values provided in this table are pseudo data for $^{13}$C$_2$D$_2$, with even parity.}
\end{deluxetable*}

\begin{figure*}[th!]
  \centering
  \includegraphics[width=0.35\textwidth]{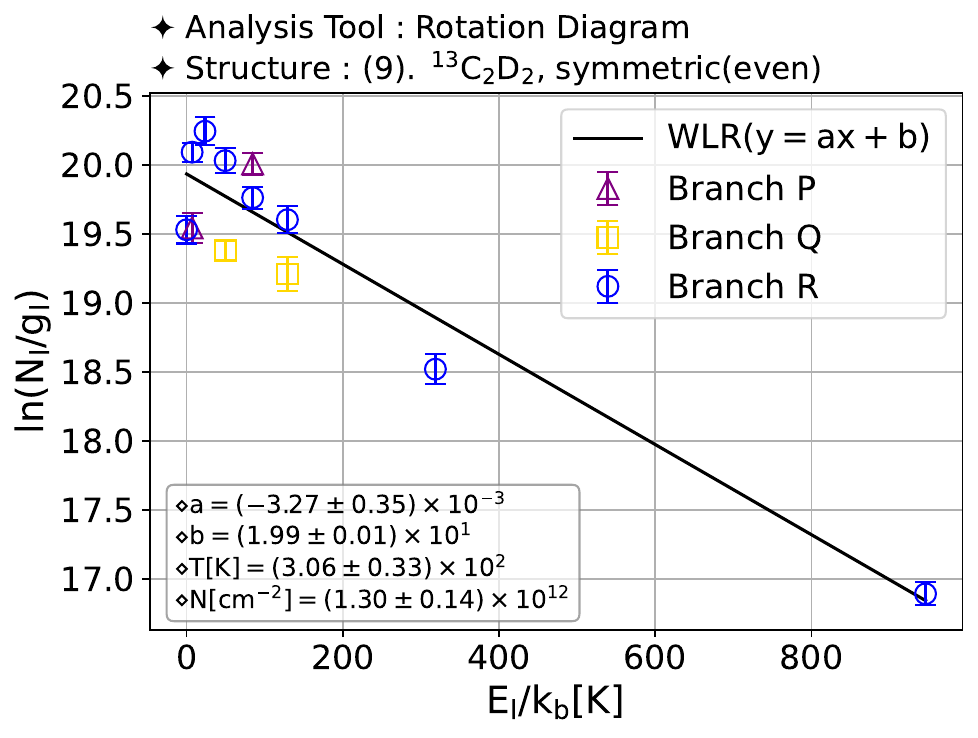}
  \includegraphics[width=0.35\textwidth]{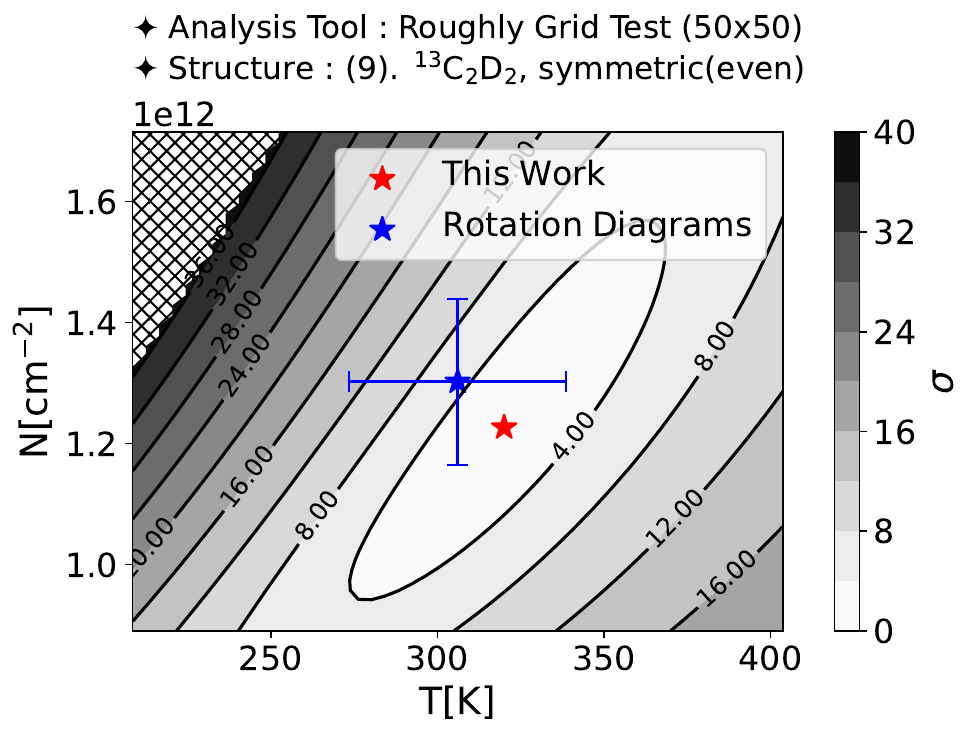}
  \includegraphics[width=0.35\textwidth]{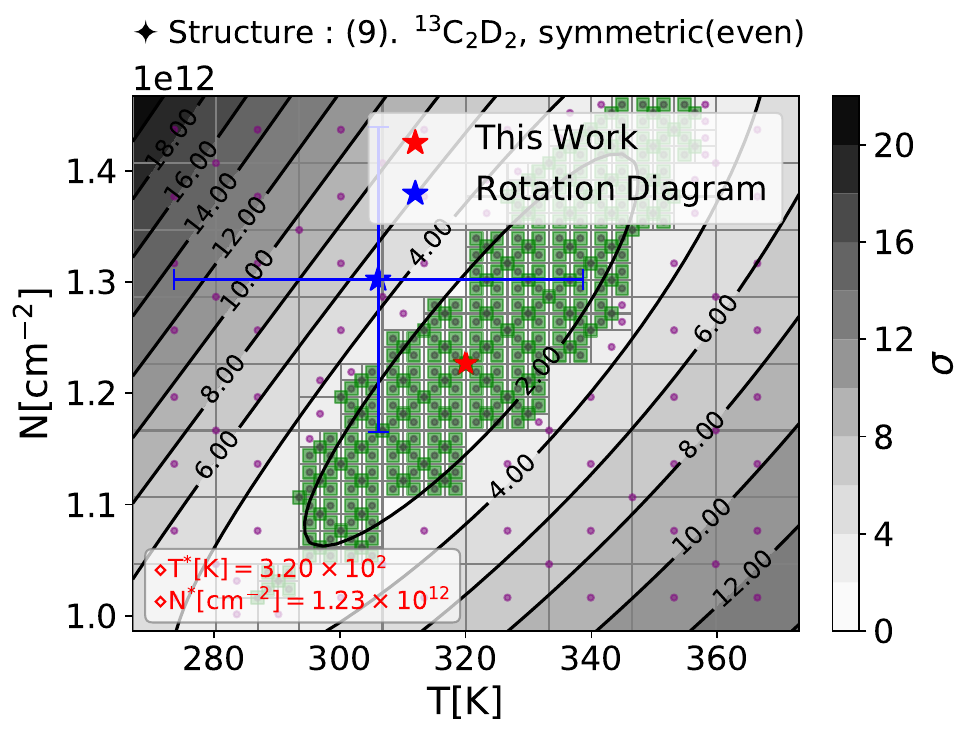}
  \includegraphics[width=0.35\textwidth]{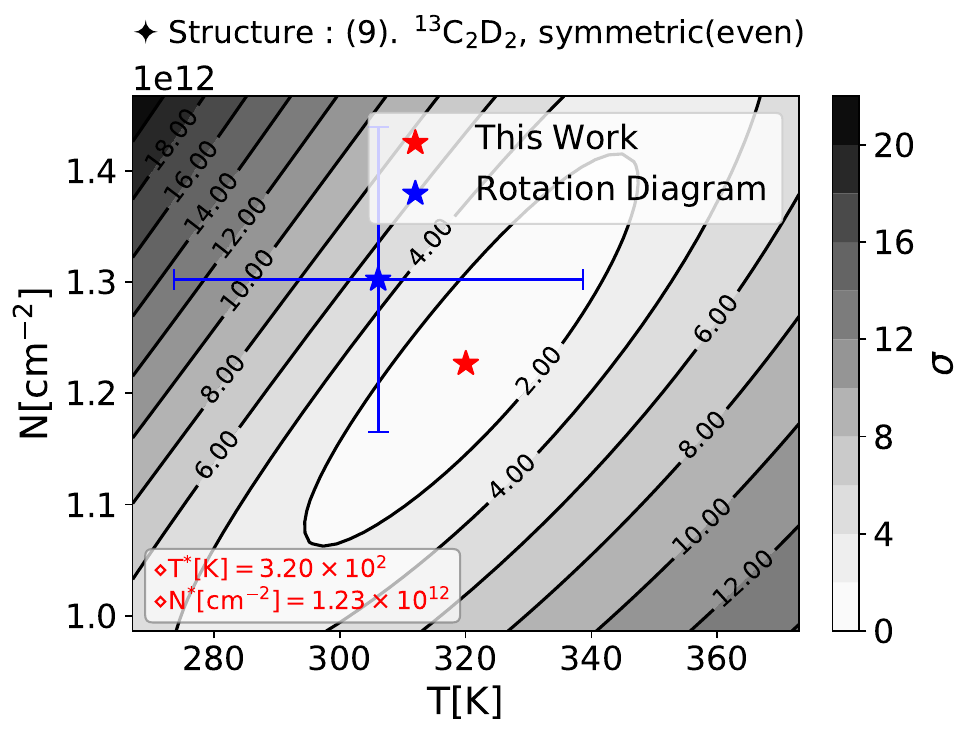}
\caption{
The TOPSEGI processes the observed values by first calculating the rotational diagram (top-left) and, based on these values, roughly identifying the minimum $\chi^2$ value in the parameter space of $T$ (temperature) and $N$ (column density) (top-right). Subsequently, the program refines the parameter space using a quadtree approach to determine the final best-fit values, $T^{*}$ and $N^{*}$ (bottom-left and bottom-right). The figure illustrates the calculation using the pseudo data of $^{13}$C$_2$D$_2$, where the parity is even, from Table \ref{tab:input_data}.
}\label{fig:TOPSEGI}
\end{figure*}

Figure \ref{fig:tips_others} illustrates the temperature dependence of the TIP (\(Q_{\text{tot}}\)) 
for all isotopologues of acetylene
over the range of 0 to 1000 K. 
Each curve represents the partition sum calculated for each isotopologue, illustrating how isotopic substitution in both hydrogen and carbon atoms influences the partition function as temperature increases.

The plot highlights that the effect of isotopic substitution becomes more pronounced with rising temperature, as the altered mass impacts vibrational energy levels, leading to differences in the partition sum. For example, isotopologues with heavier isotopes exhibit smaller gaps between vibrational energy levels, reflected in a differing partition function.

This comparison emphasizes the role of isotopic substitution in modifying the thermodynamic properties of acetylene, providing insight into how isotopic variations affect molecular stability and energy distributions at elevated temperatures.

\section{Python package : TOPSEGI}\label{apd:TOPSEGI}

Topsegi is a Python-based package developed to calculate the temperature and column density of molecules and molecular isotopologues constituting the atmospheres of planets, satellites, comets, and carbon stars, as well as interstellar medium (ISM) regions such as Orion IRc2.


The interstellar medium (ISM) refers to all types of gas, dust, and other components present between stars, primarily composed of interstellar dust and gas.

To represent ISM such as Orion IRc2, which is one of the key focuses of our research, we named the package TOPSEGI, a term from a central Korean dialect meaning "dust" or "tiny particles."

While the current version supports only acetylene and its isotopologues, future updates will include the analysis of other molecules commonly found in space, such as $\mathrm{HCN}$ and $\mathrm{^{60}C}$. 

The package integrates advanced quantum chemical calculations with robust spectral fitting techniques, offering a user-friendly framework for determining temperature ($T$) and column density ($N$) directly from observational data. It is designed to streamline the analysis process and enhance precision, particularly for isotopologues in environments with sufficiently high isotopic ratios, such as carbon-rich stars.

The package includes precomputed theoretical data for 10 isotopologues of acetylene, covering partition functions (TIPS), rotational constants, vibrational-rotational energy levels ($E_l$), degeneracy factors ($g_l$), and transition branches(see Table \ref{tab:input_data}). 
TIPS are provided for a temperature range of 1–1000 K at 1 K intervals, with values available through linear interpolation within this range. Additionally, for the P, Q, and R branches, $J_l$ are provided for values ranging from 0 to 50.
In the absence of observational data, the package provides $N/N_l$ values as a function of temperature based on theoretical data. When observational data are available, as shown in Figure \ref{fig:TOPSEGI}, the package utilizes a Quadtree algorithm to efficiently determine $T^{*}$ and $N^{*}$ values, ensuring accurate and reliable results.

This tool supports both the traditional RD method and the $\chi^2$ fitting method introduced in this study. It allows users to perform both analyses automatically and directly compare the results, highlighting the improvements achieved with the $\chi^2$ method in terms of precision and bias reduction. Furthermore, the package enables users to assess systematic biases in isotopic ratio derivations when using different methodologies, offering valuable insights into the reliability of observational results.

The package also addresses a critical gap in current spectral databases, such as HITRAN, by providing accurate theoretical data for isotopologues not included in these resources. This capability extends the applicability of the package to future observations of diverse isotopologues, enabling rapid and reliable modeling of isotopic effects. Additionally, the precomputed theoretical data included in the package ensures consistency in analyses and eliminates the need for users to perform their own quantum chemical calculations, saving time and computational resources.

The package is implemented in Python 3 and is freely available on GitHub at \texttt{https://github.com/BrownNo28/ISM}.

This package represents an important step forward in the study of interstellar chemistry and isotopic fractionation. By enabling the seamless application of the $\chi^2$ method to new datasets, it equips researchers with a powerful tool to analyze molecular spectra and explore the dynamic processes shaping astrophysical environments. Future updates to the package aim to include additional isotopologues and molecules, extending its usability to a broader range of applications in astrophysics and molecular spectroscopy. 

\bibliography{main}

\begin{thebibliography}{}
\expandafter\ifx\csname natexlab\endcsname\relax\def\natexlab#1{#1}\fi
\providecommand{\url}[1]{\href{#1}{#1}}
\providecommand{\dodoi}[1]{doi:~\href{http://doi.org/#1}{\nolinkurl{#1}}}
\providecommand{\doeprint}[1]{\href{http://ascl.net/#1}{\nolinkurl{http://ascl.net/#1}}}
\providecommand{\doarXiv}[1]{\href{https://arxiv.org/abs/#1}{\nolinkurl{https://arxiv.org/abs/#1}}}

\bibitem[{Amyay {et~al.}(2011{\natexlab{a}})Amyay, Fayt, \& Herman}]{Amway2011Fayt}
Amyay, B., Fayt, A., \& Herman, M. 2011{\natexlab{a}}, The Journal of Chemical Physics, 135, 234305, \dodoi{10.1063/1.3664626}

\bibitem[{Amyay {et~al.}(2011{\natexlab{b}})Amyay, Herman, Fayt, Campargue, \& Kassi}]{AMYAY201180}
Amyay, B., Herman, M., Fayt, A., Campargue, A., \& Kassi, S. 2011{\natexlab{b}}, Journal of Molecular Spectroscopy, 267, 80, \dodoi{https://doi.org/10.1016/j.jms.2011.02.015}

\bibitem[{Becke(1988)}]{PhysRevA.38.3098}
Becke, A.~D. 1988, Phys. Rev. A, 38, 3098, \dodoi{10.1103/PhysRevA.38.3098}

\bibitem[{{B{\'e}zard} {et~al.}(2022){B{\'e}zard}, {Vinatier}, {Greathouse}, {Giles}, {Nixon}, {Lombardo}, {Jolly}, \& {Despan}}]{bezard2022d}
{B{\'e}zard}, B., {Vinatier}, S., {Greathouse}, T., {et~al.} 2022, in European Planetary Science Congress, EPSC2022--363, \dodoi{10.5194/epsc2022-363}

\bibitem[{Brooke {et~al.}(1996)Brooke, Tokunaga, Weaver, Crovisier, Bockel{\'e}e-Morvan, \& Crisp}]{brooke1996detection}
Brooke, T., Tokunaga, A., Weaver, H., {et~al.} 1996, Nature, 383, 606, \dodoi{https://doi.org/10.1038/383606a0}

\bibitem[{Bézard {et~al.}(1991)Bézard, Romani, Conrath, \& Maguire}]{bezard1991hydrocarbons}
Bézard, B., Romani, P.~N., Conrath, B.~J., \& Maguire, W.~C. 1991, Journal of Geophysical Research: Space Physics, 96, 18961, \dodoi{https://doi.org/10.1029/91JA01930}

\bibitem[{Chubb(2018)}]{chubb2018rotation}
Chubb, K.~L. 2018, PhD thesis, UCL (University College London).
\newblock \url{https://discovery.ucl.ac.uk/id/eprint/10057098/}

\bibitem[{{de Graauw} {et~al.}(1997){de Graauw}, {Feuchtgruber}, {Bezard}, {Drossart}, {Encrenaz}, {Beintema}, {Griffin}, {Heras}, {Kessler}, {Leech}, {Lellouch}, {Morris}, {Roelfsema}, {Roos-Serote}, {Salama}, {Vandenbussche}, {Valentijn}, {Davis}, \& {Naylor}}]{de1997first}
{de Graauw}, T., {Feuchtgruber}, H., {Bezard}, B., {et~al.} 1997, \aap, 321, L13.
\newblock \url{https://ui.adsabs.harvard.edu/abs/1997A&A...321L..13D}

\bibitem[{El~Idrissi {et~al.}(1999)El~Idrissi, Liévin, Campargue, \& Herman}]{el1999vibrational}
El~Idrissi, M.~I., Liévin, J., Campargue, A., \& Herman, M. 1999, The Journal of Chemical Physics, 110, 2074, \dodoi{10.1063/1.477817}

\bibitem[{{Encrenaz} {et~al.}(1998){Encrenaz}, {Feuchtgruber}, {Atreya}, {Bezard}, {Lellouch}, {Bishop}, {Edgington}, {Degraauw}, {Griffin}, \& {Kessler}}]{encrenaz1998iso}
{Encrenaz}, T., {Feuchtgruber}, H., {Atreya}, S.~K., {et~al.} 1998, \aap, 333, L43

\bibitem[{Fayt {et~al.}(2007)Fayt, Robert, Di~Lonardo, Fusina, Tamassia, \& Herman}]{fayt2007vibration}
Fayt, A., Robert, S., Di~Lonardo, G., {et~al.} 2007, The Journal of Chemical Physics, 126, 114303, \dodoi{10.1063/1.2464101}

\bibitem[{{Fernique} {et~al.}(2015){Fernique}, {Allen}, {Boch}, {Oberto}, {Pineau}, {Durand}, {Bot}, {Cambr{\'e}sy}, {Derriere}, {Genova}, \& {Bonnarel}}]{HiPS}
{Fernique}, P., {Allen}, M.~G., {Boch}, T., {et~al.} 2015, \aap, 578, A114, \dodoi{10.1051/0004-6361/201526075}

\bibitem[{Fischer {et~al.}(2003)Fischer, Gamache, Goldman, Rothman, \& Perrin}]{FISCHER2003401}
Fischer, J., Gamache, R., Goldman, A., Rothman, L., \& Perrin, A. 2003, Journal of Quantitative Spectroscopy and Radiative Transfer, 82, 401, \dodoi{https://doi.org/10.1016/S0022-4073(03)00166-3}

\bibitem[{Fonfría {et~al.}(2008)Fonfría, Cernicharo, Richter, \& Lacy}]{fonfria2007detailed}
Fonfría, J.~P., Cernicharo, J., Richter, M.~J., \& Lacy, J.~H. 2008, The Astrophysical Journal, 673, 445, \dodoi{10.1086/523882}

\bibitem[{Foresman {et~al.}(1996)Foresman, Frisch, \& Gaussian}]{foresman1996exploring}
Foresman, J., Frisch, A., \& Gaussian, I. 1996, Exploring Chemistry with Electronic Structure Methods (Gaussian, Incorporated).
\newblock \url{https://books.google.co.kr/books?id=7ExRAAAAMAAJ}

\bibitem[{G.~Wlodarczak \& Lasne(1989)}]{wlodarczak1989rotational}
G.~Wlodarczak, J.~Demaison, J.~B., \& Lasne, M. 1989, Molecular Physics, 66, 669, \dodoi{10.1080/00268978900100411}

\bibitem[{Gamache {et~al.}(1990)Gamache, Hawkins, \& Rothman}]{GAMACHE1990205}
Gamache, R.~R., Hawkins, R.~L., \& Rothman, L.~S. 1990, Journal of Molecular Spectroscopy, 142, 205, \dodoi{https://doi.org/10.1016/0022-2852(90)90178-S}

\bibitem[{Gamache {et~al.}(2021)Gamache, Vispoel, Rey, Nikitin, Tyuterev, Egorov, Gordon, \& Boudon}]{GAMACHE2021107713}
Gamache, R.~R., Vispoel, B., Rey, M., {et~al.} 2021, Journal of Quantitative Spectroscopy and Radiative Transfer, 271, 107713, \dodoi{https://doi.org/10.1016/j.jqsrt.2021.107713}

\bibitem[{Goldsmith \& Langer(1999)}]{goldsmith1999population}
Goldsmith, P.~F., \& Langer, W.~D. 1999, The Astrophysical Journal, 517, 209, \dodoi{10.1086/307195}

\bibitem[{Gordon {et~al.}(2022)Gordon, Rothman, Hargreaves, Hashemi, Karlovets, Skinner, Conway, Hill, Kochanov, Tan, Wcisło, Finenko, Nelson, Bernath, Birk, Boudon, Campargue, Chance, Coustenis, Drouin, Flaud, Gamache, Hodges, Jacquemart, Mlawer, Nikitin, Perevalov, Rotger, Tennyson, Toon, Tran, Tyuterev, Adkins, Baker, Barbe, Canè, Császár, Dudaryonok, Egorov, Fleisher, Fleurbaey, Foltynowicz, Furtenbacher, Harrison, Hartmann, Horneman, Huang, Karman, Karns, Kassi, Kleiner, Kofman, Kwabia–Tchana, Lavrentieva, Lee, Long, Lukashevskaya, Lyulin, Makhnev, Matt, Massie, Melosso, Mikhailenko, Mondelain, Müller, Naumenko, Perrin, Polyansky, Raddaoui, Raston, Reed, Rey, Richard, Tóbiás, Sadiek, Schwenke, Starikova, Sung, Tamassia, Tashkun, {Vander Auwera}, Vasilenko, Vigasin, Villanueva, Vispoel, Wagner, Yachmenev, \& Yurchenko}]{gordon2022hitran2020}
Gordon, I., Rothman, L., Hargreaves, R., {et~al.} 2022, Journal of Quantitative Spectroscopy and Radiative Transfer, 277, 107949, \dodoi{https://doi.org/10.1016/j.jqsrt.2021.107949}

\bibitem[{Helgaker {et~al.}(2013)Helgaker, Jorgensen, \& Olsen}]{helgaker2013molecular}
Helgaker, T., Jorgensen, P., \& Olsen, J. 2013, Molecular Electronic-Structure Theory (Wiley).
\newblock \url{https://books.google.co.kr/books?id=APjLWFFxWkQC}

\bibitem[{Herman(2007)}]{herman2007acetylene}
Herman, M. 2007, Molecular Physics, 105, 2217, \dodoi{10.1080/00268970701518103}

\bibitem[{Herman {et~al.}(2003)Herman, Campargue, El~Idrissi, \& Vander~Auwera}]{Herman2003Campargue}
Herman, M., Campargue, A., El~Idrissi, M.~I., \& Vander~Auwera, J. 2003, Journal of Physical and Chemical Reference Data, 32, 921, \dodoi{10.1063/1.1531651}

\bibitem[{Herman \& Lievin(1982)}]{Herman1982acetylene}
Herman, M., \& Lievin, J. 1982, Journal of Chemical Education, 59, 17, \dodoi{10.1021/ed059p17}

\bibitem[{Herman {et~al.}(2004)Herman, Depiesse, {Di Lonardo}, Fayt, Fusina, Hurtmans, Kassi, Mollabashi, \& {Vander Auwera}}]{herman2004vibration}
Herman, M., Depiesse, C., {Di Lonardo}, G., {et~al.} 2004, Journal of Molecular Spectroscopy, 228, 499, \dodoi{https://doi.org/10.1016/j.jms.2004.05.005}

\bibitem[{Jacob {et~al.}(2020)Jacob, Menten, Wiesemeyer, G{\"u}sten, Wyrowski, \& Klein}]{jacob2020first}
Jacob, A.~M., Menten, K.~M., Wiesemeyer, H., {et~al.} 2020, Astronomy \& Astrophysics, 640, A125, \dodoi{10.1051/0004-6361/201937385}

\bibitem[{Jacquemart {et~al.}(2003)Jacquemart, Mandin, Dana, Claveau, {Vander Auwera}, Herman, Rothman, Régalia-Jarlot, \& Barbe}]{JACQUEMART2003363}
Jacquemart, D., Mandin, J.-Y., Dana, V., {et~al.} 2003, Journal of Quantitative Spectroscopy and Radiative Transfer, 82, 363, \dodoi{https://doi.org/10.1016/S0022-4073(03)00163-8}

\bibitem[{Karakas(2010)}]{karakas2010updated}
Karakas, A.~I. 2010, Monthly Notices of the Royal Astronomical Society, 403, 1413, \dodoi{10.1111/j.1365-2966.2009.16198.x}

\bibitem[{{Lambert} {et~al.}(1986){Lambert}, {Gustafsson}, {Eriksson}, \& {Hinkle}}]{lambert1986chemical}
{Lambert}, D.~L., {Gustafsson}, B., {Eriksson}, K., \& {Hinkle}, K.~H. 1986, \apjs, 62, 373, \dodoi{10.1086/191145}

\bibitem[{Laraia {et~al.}(2011)Laraia, Gamache, Lamouroux, Gordon, \& Rothman}]{LARAIA2011391}
Laraia, A.~L., Gamache, R.~R., Lamouroux, J., Gordon, I.~E., \& Rothman, L.~S. 2011, Icarus, 215, 391, \dodoi{https://doi.org/10.1016/j.icarus.2011.06.004}

\bibitem[{Lee {et~al.}(1988)Lee, Yang, \& Parr}]{PhysRevB.37.785}
Lee, C., Yang, W., \& Parr, R.~G. 1988, Phys. Rev. B, 37, 785, \dodoi{10.1103/PhysRevB.37.785}

\bibitem[{McCarthy {et~al.}(2019)McCarthy, Gottlieb, \& Cernicharo}]{mccarthy2019building}
McCarthy, M.~C., Gottlieb, C.~A., \& Cernicharo, J. 2019, Journal of Molecular Spectroscopy, 356, 7, \dodoi{https://doi.org/10.1016/j.jms.2018.11.018}

\bibitem[{McQuarrie(2008)}]{mcquarrie2008quantum}
McQuarrie, D. 2008, Quantum Chemistry, G - Reference,Information and Interdisciplinary Subjects Series (University Science Books).
\newblock \url{https://books.google.co.kr/books?id=zzxLTIljQB4C}

\bibitem[{M\o{}ller \& Plesset(1934)}]{PhysRev.46.618}
M\o{}ller, C., \& Plesset, M.~S. 1934, Phys. Rev., 46, 618, \dodoi{10.1103/PhysRev.46.618}

\bibitem[{Nickerson {et~al.}(2023)Nickerson, Rangwala, Colgan, DeWitt, Monzon, Huang, Acharyya, Drozdovskaya, Fortenberry, Herbst, \& Lee}]{nickerson2023mid}
Nickerson, S., Rangwala, N., Colgan, S. W.~J., {et~al.} 2023, The Astrophysical Journal, 945, 26, \dodoi{10.3847/1538-4357/aca6e8}

\bibitem[{Nixon {et~al.}(2007)Nixon, Achterberg, Conrath, Irwin, Teanby, Fouchet, Parrish, Romani, Abbas, LeClair, Strobel, Simon-Miller, Jennings, Flasar, \& Kunde}]{nixon2007meridional}
Nixon, C., Achterberg, R., Conrath, B., {et~al.} 2007, Icarus, 188, 47, \dodoi{https://doi.org/10.1016/j.icarus.2006.11.016}

\bibitem[{O\&apos;Dell(2001)}]{o2001orion}
O\&apos;Dell, C.~R. 2001, Annual Review of Astronomy and Astrophysics, 39, 99, \dodoi{https://doi.org/10.1146/annurev.astro.39.1.99}

\bibitem[{Pagel(2009)}]{pagel2009nucleosynthesis}
Pagel, B. E.~J. 2009, Nucleosynthesis and chemical evolution of galaxies (Cambridge University Press), \dodoi{https://doi.org/10.1017/CBO9780511812170}

\bibitem[{Pentsak {et~al.}(2024)Pentsak, Murga, \& Ananikov}]{pentsak2024role}
Pentsak, E.~O., Murga, M.~S., \& Ananikov, V.~P. 2024, ACS Earth and Space Chemistry, 8, 798, \dodoi{10.1021/acsearthspacechem.3c00223}

\bibitem[{Perdew {et~al.}(1996)Perdew, Burke, \& Ernzerhof}]{PhysRevLett.77.3865}
Perdew, J.~P., Burke, K., \& Ernzerhof, M. 1996, Phys. Rev. Lett., 77, 3865, \dodoi{10.1103/PhysRevLett.77.3865}

\bibitem[{Rangwala {et~al.}(2018)Rangwala, Colgan, Gal, Acharyya, Huang, Lee, Herbst, deWitt, Richter, Boogert, \& McKelvey}]{rangwala2018high}
Rangwala, N., Colgan, S. W.~J., Gal, R.~L., {et~al.} 2018, The Astrophysical Journal, 856, 9, \dodoi{10.3847/1538-4357/aaab66}

\bibitem[{Robert {et~al.}(2009)Robert, Amyay, Fayt, Di~Lonardo, Fusina, Tamassia, \& Herman}]{Robert2007}
Robert, S., Amyay, B., Fayt, A., {et~al.} 2009, The Journal of Physical Chemistry A, 113, 13251, \dodoi{10.1021/jp904000q}

\bibitem[{Stahler \& Palla(2008)}]{stahler2008formation}
Stahler, S., \& Palla, F. 2008, The Formation of Stars (Wiley).
\newblock \url{https://books.google.co.kr/books?id=X91UBLr64FMC}

\bibitem[{Woods(2010)}]{woods2009carbon}
Woods, P.~M. 2010, Carbon isotope measurements in the Solar System.
\newblock \doarXiv{0901.4513}

\end{thebibliography}

\bibliographystyle{aasjournal}

\end{document}